\documentclass [10pt,twocolumn]{IEEEtran}
\usepackage{algorithm,algorithmic,amsbsy,amsmath,amssymb,epsfig,bbm,mathrsfs,multirow,amsthm}
\usepackage[T1]{fontenc}
\usepackage[latin9]{inputenc}
\usepackage{amsmath}
\usepackage{fixltx2e}
\usepackage{algorithm}
\usepackage{array,multirow,graphicx}
\usepackage{color}
\usepackage{cite}
\usepackage{float}
\usepackage{makecell}
\usepackage[x11names,dvipsnames,table]{xcolor} 
\usepackage{colortbl}
\usepackage{multirow}
\definecolor{mygray}{gray}{0.6}
\definecolor{myblue}{rgb}{0.8,0.85,1} 
\usepackage{array}
\newcolumntype{L}[1]{>{\raggedright\let\newline\\\arraybackslash\hspace{0pt}}m{#1}}
\newcolumntype{C}[1]{>{\centering\let\newline\\\arraybackslash\hspace{0pt}}m{#1}}
\newcolumntype{R}[1]{>{\raggedleft\let\newline\\\arraybackslash\hspace{0pt}}m{#1}}
\usepackage{soul}

\usepackage{array, tabularx, boldline}
\usepackage{graphicx}
\usepackage{cellspace}
\begin{document}
\title{Applications of Economic and Pricing Models for Wireless Network Security: A Survey}

\author{Nguyen Cong Luong, Dinh Thai Hoang,  Ping Wang, \textit{Senior Member, IEEE}, Dusit Niyato, \textit{Fellow, IEEE}, and Zhu Han, \textit{Fellow, IEEE}
\thanks{N.~C.~Luong, D.~T.~Hoang, P.~Wang, and D.~Niyato are with School of Computer Science and Engineering, Nanyang Technological University, Singapore. E-mails: clnguyen@ntu.edu.sg, thdinh@ntu.edu.sg, wangping@ntu.edu.sg, dniyato@ntu.edu.sg.}
\thanks{Z.~Han is with Electrical and Computer Engineering, University of Houston, Houston, TX, USA, and also with the Department of Computer Science and Engineering, Kyung Hee University, Seoul, South Korea. E-mail: zhan2@uh.edu.} 
}

\maketitle
\begin{abstract}  
This paper provides a comprehensive literature review on applications of economic and pricing theory to security issues in wireless networks. Unlike wireline networks, the broadcast nature and the highly dynamic change of network environments pose a number of nontrivial challenges to security design in wireless networks. While the security issues have not been completely solved by traditional or system-based solutions, economic and pricing models recently were employed as one efficient solution to discourage attackers and prevent attacks to be performed. In this paper, we review economic and pricing approaches proposed to address major security issues in wireless networks including eavesdropping attack, Denial-of-Service (DoS) attack such as jamming and Distributed DoS (DDoS), and illegitimate behaviors of malicious users. Additionally, we discuss integrating economic and pricing models with cryptography methods to reduce information privacy leakage as well as to guarantee the confidentiality and integrity of information in wireless networks. Finally, we highlight important challenges, open issues and future research directions of applying economic and pricing models to wireless security issues.

\end{abstract}

\section{Introduction}
\label{sec:intro}

Wireless networks have been widely used in several applications in different areas. The shared and easy to access wireless medium is undoubtedly the biggest advantage of wireless networks. However, the exposed nature of the wireless medium makes wireless networks vulnerable to various attacks. The most commonly seen attack in wireless networks is eavesdropping in which attackers aim at acquiring important/private information of users. Another well-known attack is Denial-of-Service (DoS) such as jamming attack and Distributed DoS (DDoS) attack which attempt to interfere and disrupt network operations by exhausting the resources available to legitimate systems and users. Although the attacks have different strategies and objectives, they all lead to serious consequences such as degrading the network performance and Quality of Service (QoS) as well as losing important data, reputations, and revenue.

To combat the attacks, several security mechanisms have been proposed. For example, for the eavesdropping attack, cryptographic techniques \cite{delfs2007symmetric, salomaa2013public} have been commonly applied at the upper network layers to reduce the attacker's ability to decode private information. For the DoS attack, frequency hopping techniques \cite{popovski2006strategies,popper2010anti} have been adopted as anti-jamming solutions. However, such traditional techniques face many challenges or even may not work in future generation wireless networks that become more decentralized and ad-hoc in nature. For example, cryptographic techniques often require centralized authorities, additional secure channels for key exchanges, and more computation power for encryption/decryption. The requirements are hardly fulfilled in decentralized mobile environments. 

\subsection{What and Why Economic and Pricing Approaches}
\label{what_why_economic}

Economic and pricing methods have been recently developed to address various security issues in wireless networks. They allow to model and analyze modern distributed system operations such as negotiations among independent and selfish users. Thus they can be adopted in the wireless security scenarios to model and analyze complex interactions among attackers and legitimate users/defenders. Through the interactions, the legitimate users can learn or predict the malicious behaviors of the attackers, then having optimal defending and reaction strategies based on equilibrium analysis. Besides, economic and pricing models provide incentive mechanisms to prevent the attackers from launching attacks or executing illegitimate behaviors. In other words, the traditional or system-based approaches are considered to be
the "hard" approaches that require additional hardware and software implementations while the economic and pricing approaches are considered to be the "soft" approaches that motivate and incentivize the users not to launch attack actions as it is not profitable for them to do so. The economic and pricing approaches are natural and inherent in the decision making of all the network entities, thus more efficient and more effective to employ.

\begin{figure}[t!]
 \centering
\includegraphics[width=5 cm, height=5cm]{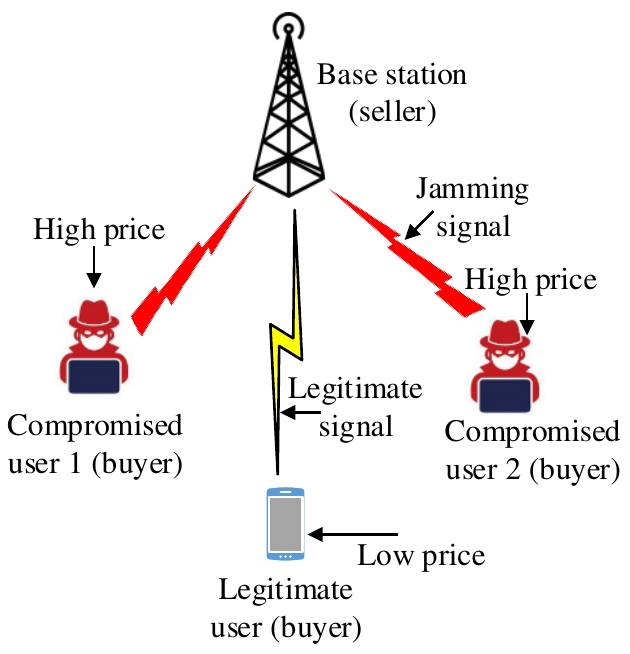}
 \caption{DDoS attack prevention using Bayesian-optimal pricing \cite{chorppathbayesian}.}
 \label{DDoS_attack_for_intro}
\end{figure}

Fig.~\ref{DDoS_attack_for_intro} illustrates an example of using economic and pricing models to prevent the DDoS attack in wireless networks. For the DDoS attack~\cite{zargar2013survey}, an external attacker controls a large number of \textit{compromised} users inside the network to transmit radio jamming signals with high power to a target system, and makes the system unavailable to respond to any service requests from the legitimate users. Consider the simple model in Fig.~\ref{DDoS_attack_for_intro} which has two compromised users 1 and 2, the legitimate user (target of the attackers), and the base station. Note that the compromised users and the legitimate user are buyers while the base station acts as a resource seller. In particular, the compromised users are assumed to be rational. Since the compromised users need to consume much network resources for their attack actions, they are subject to the resource prices. Therefore, setting high resource prices to the compromised users discourages them to spend more powers to perform the attack actions. The Bayesian-optimal pricing model \cite{chorppathbayesian} can be adopted to easily achieve the purpose. Specifically, using the pricing model, the resource price for each user is proportional to the probability that the user performs a malicious action. Since the compromised users have high probabilities of performing malicious actions, the resource prices for them are high. On the contrary, low prices are set for the legitimate users. The users are demotivated to harm or attack other users because of the high cost. As shown in the simulation results in \cite{chorppathbayesian}, the additional cost for a malicious user significantly decreases when the degree of maliciousness of the user decreases.

The example discussed above shows that the DDoS attack can be simply addressed by pricing without requiring additional hardware and software. For wireless attacks and security issues in general, the economic and pricing approaches provide the following advantages:
\begin{itemize}
\item Economic and pricing models as distributed solutions maximize the secrecy capacity for the wireless communications to combat the eavesdropping attack. These distributed solutions do not require additional secure channels for key exchanges as well as the complete knowledge of all channel information. However, these solutions often need the help of friendly jammers. The specific approaches are reviewed in Section~\ref{sec:Survey_eavesdropper_attack}. 

\item Attackers can launch active attacks to cause the interference to the channels. Economic and pricing models allow legitimate users with good channels to trade their capacities with other legitimate users under interfered channels in order to maximize the total utility of all legitimate users in the network. More details are presented in Section \ref{sec:Survey_DoS_attack}. 


\item As shown in the above example, an attacker can launch its attacks through a large number of compromised users inside the network. The compromised users are typically distributed across network. Pricing models as decentralized solutions are efficiently used to combat them. The reviewed approaches are given in Section \ref{sec:Survey_DoS_attack}.

\item The distributed nature of the compromised users makes them extremely difficult to be detected and traced. Economic and pricing models provide the legitimate users sufficient incentive to perform monitoring of other users to quickly detect and isolate the compromised users. Section~\ref{sec:Survey_DoS_attack} discusses further the related approaches.

\item In realistic wireless environments, the legitimate users often face the risks of physical attacks. The risks arise if their location information is revealed. By combining with encryption methods, economic and pricing models provide privacy preserving mechanisms to solve the leakage of users' location information. Moreover, the economic and pricing models provide incentive mechanisms to users to form an \textit{anonymity} set so that the attackers cannot identify a particular user within the anonymity set. Section~\ref{sec:Survey_privacy_concerns} presents the specific approaches. 

\item Apart from the aforementioned issues, there are other security issues in spectrum allocation process in wireless networks. The security issues mainly arise from the cheating and illegitimate behaviors of users which severely deteriorate the efficiency of the allocation. Economic and pricing models are used to guarantee that the users cannot increase their payoffs with such illegal behaviors. The related work is reviewed in Section \ref{sec:Survey_collusion_false}.
\end{itemize}

Nevertheless, a common limitation of the economic and pricing approaches is that they require stakeholders in the system to be rational and react to the strategy that the stakeholders achieve the highest benefit, whereas in some cases this assumption does not hold. 
\subsection{Contributions of the Paper}
\label{Contribution_Paper}

Although there are several surveys related to the wireless security, they only focus on a particular wireless network and non-economic approaches. For example, surveys of security issues in Wireless Sensor Networks (WSNs) are given in \cite{wang2006survey} and \cite{pathan2006security}. A survey of techniques against DDoS attacks in Mobile Ad-hoc Network (MANETs) is presented in \cite{zargar2013survey}. There are also surveys or tutorials related to the general wireless security, but they only consider one specific security issue, e.g., anti-jamming techniques \cite{vadlamani2016jamming}, and physical layer security \cite{mukherjee2014principles}. Recent surveys discuss the applications of game theory for the security in computer-communication networks \cite{manshaei2013game}, \cite{liang2013game}. However, they are not specifically from economic and pricing perspectives, which are emerging as a promising approach. To the best of our knowledge, there is no survey specifically discussing the applications of economic and pricing models to address the security issues in wireless networks. This motivates us to deliver the survey with the comprehensive literature review on economic and pricing models developed for the wireless security.

\begin{figure*}[ht]
\centering
\includegraphics[width=\linewidth]{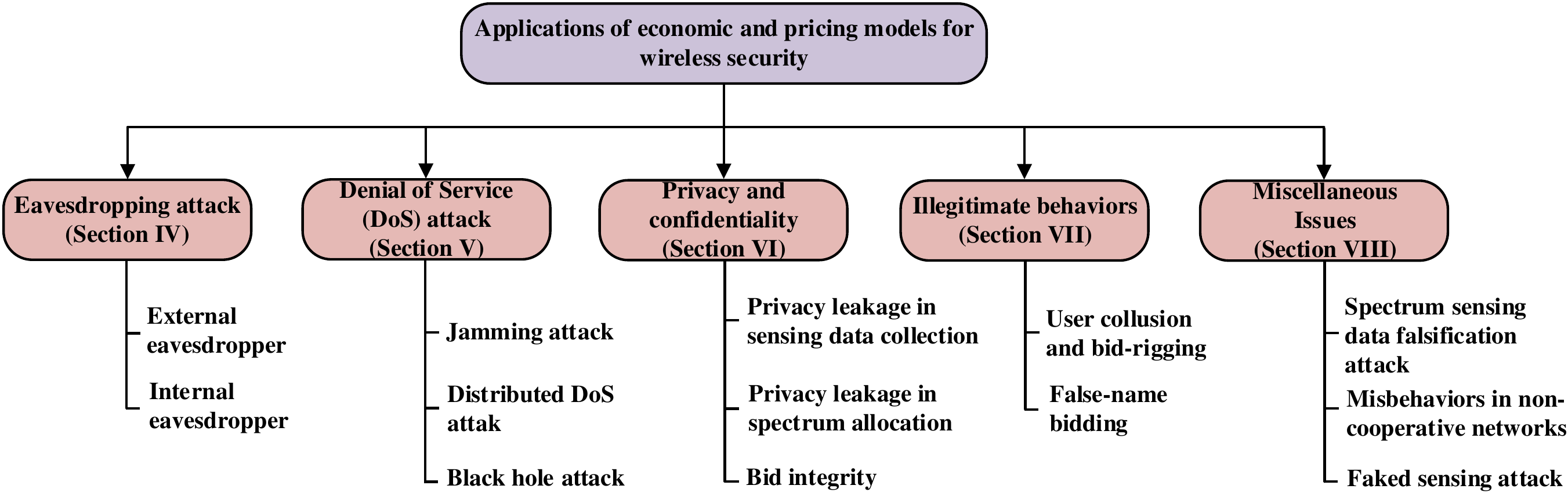}
 \caption{A taxonomy of the applications of economic and pricing models for wireless security.}
 \label{Application_pricing_model}
\end{figure*}

For convenience, the related works in this survey are categorized according to wireless security issues which have been addressed by economic and pricing models as shown in Fig~\ref{Application_pricing_model}. Specifically, the economic and pricing models have solved the following major security issues:  
\begin{itemize}
\item \textit{Eavesdropping attacks} with the common assumption that there are supports from friendly jammers in the system.
\item \textit{DoS attacks} with the common assumption that stakeholders in the system are rational and react to the strategy that they achieve the highest benefit.
\item \textit{Privacy and confidentiality} with the common assumption that there is no collusion among stakeholders in the system.
\item \textit{Illegitimate behaviors} with the common assumption that stakeholders in the system may have cheating behaviors to maximize their own utilities, but they do not cause damage to other users. 
\end{itemize}
 Additionally, some other security issues such as the spectrum sensing data falsification attack and faked sensing attack are also discussed. Main results, advantages, drawbacks, and the future directions of each approach are highlighted.

The remainder of this paper is organized as follows. Section~\ref{sec:Intro_Security} gives an overview of security issues in wireless networks. The basics and fundamentals of the economic and pricing models used to address the wireless security issues are given in Section~\ref{sec:Intro_Price}. Section~\ref{sec:Survey_eavesdropper_attack} reviews the applications of the economic and pricing models for securing wireless communication against eavesdropping attacks. Section \ref{sec:Survey_DoS_attack} discusses the economic and pricing approaches for the DoS attack. Section \ref{sec:Survey_privacy_concerns} presents economic and pricing approaches for the information security concerning the information privacy and confidentiality. Sections \ref{sec:Survey_collusion_false}  considers how to use the pricing strategies to deal with illegitimate and cheating behaviors in wireless networks. Furthermore, applications of economic and pricing models for the spectrum sensing data falsification and faked sensing attacks are given in Section \ref{sec:Survey_security_issues}. Section~\ref{sec:Open_issues} summarizes important challenges, open issues, and future research directions. Finally, the paper is concluded in Section~\ref{sec:Conclusion}. The list of abbreviations used in this paper is given in Table~\ref{tab:table_abb}. 

\begin{table}[h!]
\scriptsize
  \caption{Major abbreviations}
  \label{tab:table_abb}\centering
  \begin{tabularx}{8.7cm}{|Sl|X|}
    \hline
  \cellcolor{mygray} \textbf{Abbreviation} &   \cellcolor{mygray} \textbf{Description} \\   
    \hline
  ACA-A& Alternative Ascending Clock Auction \\ 
         \hline
  ACA-T& Traditional Ascending Clock Auction \\ 
       \hline
   CA/CDF &Cryptographic Authority/Cumulative Distribution Function\\  
       \hline
   CIA &Confidentiality, Integrity and Availability triad\\  
      \hline
   CRN  & Cognitive Ratio Network\\   
         \hline
   DoS/DDoS & Denial-of-Service/Distributed Denial-of-Service\\
     \hline
EBV&Encrypted Bit Vector\\
\hline
   FJ   & Friendly Jamming\\
     \hline
  HMAC&Hash Message Authentication Code\\
        \hline
KKT& Karush-Kuhn-Tucker \\
         \hline
LSA& Licensed Shared Access\\
\hline
   MANET& Mobile Ad hoc NETwork \\
      \hline
    NUM  & Network Utility Maximization\\
        \hline
   OPE & Order Preserving Encryption\\
       \hline
 PD/PU/SU&Primary Destination/Primary User/Secondary User\\
        \hline
 SINR   &Signal-to-Interference-plus-Noise Ratio \\
       \hline
 SSDF& Spectrum Sensing Data Falsification\\
      \hline
TLC/TTP & Time Lapse Cryptography/Trusted Third Party\\
      \hline
VCG/GSP & Vickrey-Clarke-Groves/Generalized Second-Price auction \\
        \hline
  \end{tabularx}
\end{table}

\section{Overview of wireless security issues}
\label{sec:Intro_Security}
Wireless networks play an extremely important role in many applications. However, security for wireless networks remains a big challenge. The main reason is that accessing wireless networks from users does not require physical connections. It is thus much easier for attackers to launch various attacks to target users or systems in wireless networks. This section briefly reviews security issues in wireless networks which have been addressed by economic and pricing models. Specifically, we first introduce different types of attackers and users in wireless networks. Then, we discuss two major attacks in wireless networks, i.e., eavesdropping and DoS. Finally, we define some information security issues and malicious behaviors of users in wireless networks. 

\subsection{Users and attackers in wireless networks}
\label{sec:Intro_type_of_node}
Different types of users\footnote{Users and nodes are used interchangeably in the rest of the paper.} and attackers as well as their key features in typical wireless networks are shown in Table~\ref{tab:table_type_user}. 

\begin{table}[h!]
\scriptsize
  \caption{Types of users/attackers and their features}
  \label{tab:table_type_user}\centering
 \begin{tabularx}{8.9cm}{|L{1.7cm}|X|}
    \hline
  \cellcolor{mygray} \textbf{Users/attackers} &   \cellcolor{mygray} \textbf{Key features} \\ 
    \hline
Altruistic users& Regular users or legitimate users which behave in such a way to improve the overall network performance \\ 
         \hline
Selfish users & Regular users or legitimate users which behave in such a way to maximize their own utilities\\ 
         \hline
Compromised users/bots&  Users within a network which launch insider attacks to a target in the network \\ 
       \hline
Malicious users& Adversarial users which attempt to damage or harm other users or systems\\  
       \hline
Eavesdropper/wiretapper &Unauthorized receiver which performs illegally capturing and reading data packets from legitimate sources\\  
      \hline
Jammer  & Adversarial user which launches the DoS attack by transmitting a high power noise signal to degrade or corrupt the signal at the intended receiver\\   
         \hline
Friendly jammer & Assist legitimate sources by transmitting a so-called \textit{Friendly Jamming} (FJ) signal to reduce the data rate that is leaking from the source to the eavesdropper\\
     \hline
Active eavesdropper   & Adversarial user which performs both jamming and eavesdropping\\
     \hline
  \end{tabularx}
\end{table}

\subsection{Eavesdropping Attack}
\label{sec:Intro_Security_Physical_layer}
Due to the broadcast nature of wireless environments, an eavesdropper is able to capture information from legitimate communications. There are several cryptographic techniques which have been implemented at the upper layers of wireless networks, e.g., a secret key cryptography \cite{delfs2007symmetric} and public key cryptography \cite{salomaa2013public}, to reduce the eavesdropper's ability to decode the private information. However, these techniques often require additional secure channels for key exchanges and more computation power for encryption/decryption which are both scare and expensive in mobile environments. Thus security algorithms on the physical layer have been proposed as an alternative. Exploiting physical characteristics of wireless channels, e.g., channel gains, to protect the wireless communication against eavesdropping attacks is called \textit{physical layer security} \cite{zou2015improving}. 
\begin{figure}[t!]
 \centering
\includegraphics[width=6.1cm, height=3.8cm]{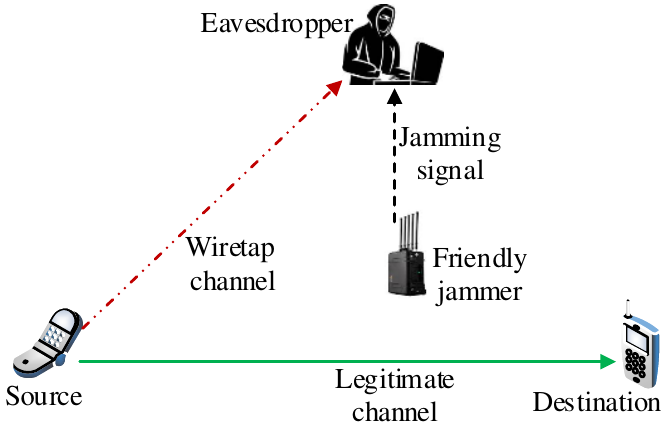}
 \caption{Eavesdropping attack model with a source, a destination, an eavesdropper, and a friendly jammer.}
 \label{Eavesdropper_Attack}
\end{figure}
Its objective is to maximize the transmission rate of reliable information transmitted from a legitimate source to an intended destination at which the eavesdropper is unable to decode the information. The maximum reliable rate is namely \textit{secrecy capacity} or \textit{secrecy rate}. It is defined as the difference between the capacity of the source-destination channel, i.e., the legitimate channel, and that of the source-eavesdropper channel, i.e., the wiretap channel \cite{dong2010improving}. 

To improve the secrecy capacity, one common solution is to employ Friendly Jamming (FJ) signal to degrade the signal quality at the eavesdropper as shown in Fig.~\ref{Eavesdropper_Attack}. However, employing the FJ signal also decreases the quality of the legitimate channel. This problem can be considered to be a power allocation issue which has been efficiently solved by pricing models such as auctions or game theory. Therefore, it is interesting to discuss how to improve the secrecy capacity by using economic and pricing models which are reviewed in Section \ref{sec:Survey_eavesdropper_attack}.

\subsection{Denial-of-Service (DoS) attack}
\label{sec:Intro_DoS_attack}
A DoS attack is an active attack in which adversaries attempt to exhaust resources available to legitimate users. There are three common types of the DoS attack in wireless networks, i.e., the jamming attack, DDoS attack, and black hole attack.

\subsubsection{Jamming attack}
\label{sec:Intro_DoS_attack_jamming}
In the jamming attack, adversaries as jammers transmit Radio Frequency (RF) jamming signals with high power to disrupt or cause interference to the legitimate communications between sources and destinations. While attempting to cause the interference, the jammers are subject to the cost of power. Anti-jamming schemes using power pricing strategies \cite{zhang2014cooperative}, \cite{tang2016combating} can be applied to discourage jammers to consume power for their jamming actions.  

\subsubsection{Distributed DoS (DDoS) attack}
\label{sec:Intro_DoS_attack_Distributed}
A DDoS attack is a large-scale, coordinated attack on the availability of network resources. This attack is typically launched indirectly through a large number of compromised nodes, i.e., bots, inside the network. The bots are illegally tampered and manipulated to launch the DDoS attack to a target system, e.g., a base station. They are practically controlled by an external attacker of the network. The DDoS attack is known as one of the most severe attacks due to the use of thousands of bots distributed over the network. However, since the bots consume network resources, they can be effectively prevented through behavior-based pricing policies which are reviewed in Section \ref{sec:Survey_DoS_attack_Botnet}.

\subsubsection{Black hole attack}
\label{sec:Intro_DoS_attack_Black}
The black hole attack attemps to destruct network services such as packet routing or forwarding. Indeed, in a typical routing protocol, a source broadcasts a \textit{route request} packet to all intermediate nodes before sending data packets to its destination. Then, the nodes which are in the route towards the destination reply the source with \textit{route reply} packets. There may exist a malicious node which claims itself of having the shortest route to the destination and then later drops incoming packets instead of forwarding them to the destination. Such an attack is called \textit{black hole attack}, and the malicious node which performs the attack is called \textit{black hole node}. Pricing mechanisms are applied to provide nodes incentives to forward packets rather than dropping them \cite{agah2006security}. 

\subsection{Information Security Issues}
\label{sec:Intro_Security_privacy}

Information security is the prevention of unauthorized access, damage, leakage, modification, and recording of information \cite{joshi2001security}. Generally, the information security is defined as the CIA attributes including confidentiality, integrity, and availability. In particular, the confidentiality makes the information unavailable or undisclosed to unauthorized users while the integrity guarantees that the information is not modified or destroyed in an unauthorized manner. Besides, information privacy is perceived as part of information security. 

 Information privacy is defined as the ability of the user to control one's personal information when this information is acquired and used \cite{stone1983field}. Enhancing the information privacy of users requires \textit{privacy preserving} mechanisms. Traditional solutions in wireless networks are to employ \textit{ciphertext}. Ciphertext is the result of encryption performed on \textit{plaintext} using an algorithm, called a \textit{cipher} or \textit{cypher} \cite{stinson2005cryptography}. Therefore, the ciphertext is known as encrypted information which contains a form of the original \textit{plaintext} that is unreadable without the proper cipher to decrypt it. The key used in the encryption cipher is called a \textit{cryptographic key}. The ciphertext can be generated by cryptographic algorithms such as Time Lapse Cryptography (TLC) \cite{rabin2006time} and keyed-Hash Message Authentication Code (HMAC) \cite{bellare1996keying}. The TLC uses a private key to encrypt each user's information, and the private key is not known to anyone until a predefined future time. After that time, the private key is published, and anyone can decrypt the ciphertext. The HMAC is a hashing method which uses a secret cryptographic key in conjunction with a cryptographic hash function, e.g., MD5 or SHA-1 \cite{bellare1996keying}. 

Besides the TLC and HMAC, users' information can be encrypted by the Paillier cryptosystem \cite{paillier1999public}, i.e.,  a probabilistic asymmetric algorithm for public key cryptography, or by the Order Preserving Encryption (OPE) \cite{agrawal2004order}, i.e., a deterministic encryption scheme which encrypts plaintexts to ciphertexts with the same order. An interesting property of these schemes is \textit{homomorphic} which allows computations to be performed on ciphertexts, e.g., encrypted bids, and the decrypted results match those of operations performed on the plaintext, e.g., original bids.

\subsection{Illegitimate Behaviors In Wireless Networks}
\label{sec:Cheating_behaviors}
Economic and pricing approaches have recently been applied to wireless networks with an increasing trend. In particular, they have been used to solve several wireless security issues as mentioned in Section \ref{sec:intro}. However, introducing pricing to wireless systems can lead to security issues that deserve discussion in our survey. These security issues mostly come from using auctions in which users as buyers, i.e., bidders, submit their bids, i.e., bidding prices, for the requested resources, e.g., spectrum. To increase payoffs, the users may perform the illegitimate behaviors as described in Table~\ref{tab:table_illegitimate_beha}. 
\begin{table}[h!]
\scriptsize
  \caption{Illegitimate behaviors in wireless networks}
  \label{tab:table_illegitimate_beha}\centering
\begin{tabularx}{8.7cm}{|L{1.3cm}|X|X|}
    \hline
  \cellcolor{mygray} \textbf{Behaviors} &   \cellcolor{mygray} \textbf{Definitions} &\cellcolor{mygray} \textbf{Consequences}\\ 
    \hline
False-name bid cheating&Dishonest action in which multiple names, i.e., e-mail addresses, are submitted by a single user&  Reduce the seller's revenue and other bidders' utilities \\
    \hline
Collusion& Cheating behavior in which a bidder as a buyer seeks out and persuades other bidders to propose a lower price to the seller.  A collection of the bidders is called a \textit{bidding ring}& Reduce the seller's revenue and pose severe threats to the efficiency of spectrum allocation \\ 
    \hline
Bid-rigging& A form of collusion in which an
untrustworthy auctioneer conspires with greedy bidders& Illegally fix the price and manipulate auction\\ 
     \hline
  \end{tabularx}
\end{table}

\textbf{Summary:} This section describes security issues in wireless networks involving eavesdropping attack, DoS attack, and misbehaviors of users in wireless networks. The common challenge of the security issues is to determine exactly locations of attackers, malicious users as well as their misbehaviors since the attackers and malicious users may be distributed across over the network. Economic and pricing models can enhance secrecy capacity and encourage malicious users to refrain from misbehaving without requiring the perfect knowledge of their locations. To easily understand how to apply the economic and pricing models for the security issues in wireless networks, the next section presents basics and fundamentals of economic and pricing models used in this survey. 

\section{Overview and fundamentals of economic and pricing theory for security in wireless networks}
\label{sec:Intro_Price}
\begin{figure*}[ht]
 \centering
\includegraphics[width=\linewidth]{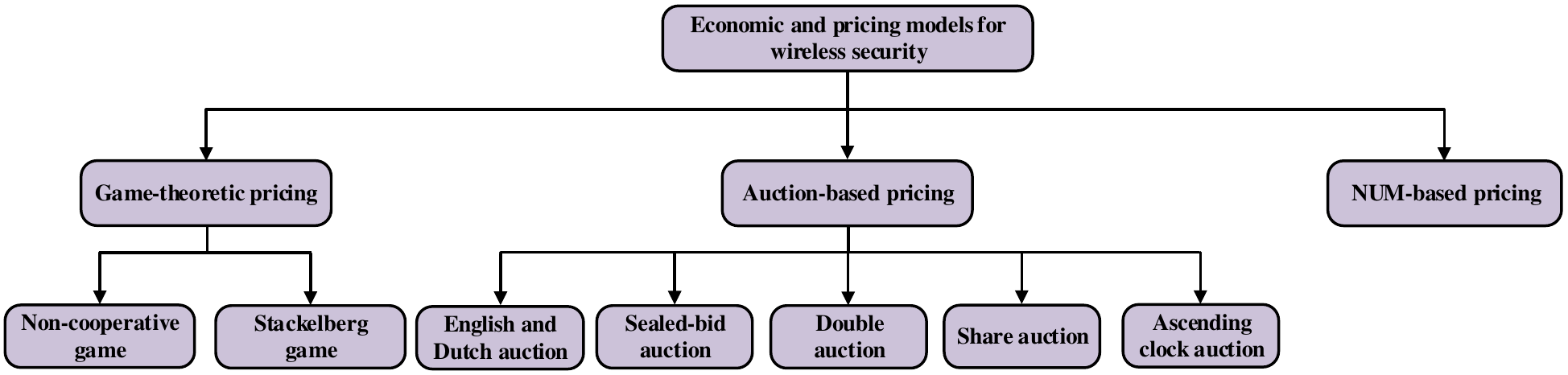}
 \caption{A taxonomy of economic and pricing models for wireless security.}
 \label{taxonomy_pricing_model}
\end{figure*}

Economic and pricing approaches have been recently applied to address various security issues in wireless networks due to the aforementioned benefits. In this section, we present the background of the economic and pricing models as well as the rationale behind their use to enhance the wireless security. A taxonomy of economic and pricing models used for the wireless security is shown in Fig.~\ref{taxonomy_pricing_model}. Note that there may exist several different pricing models depending on how to set the price as presented in \cite{luong2016data} and \cite{nguyen2017resource}. For example, when the price is set based on cost analysis, we have cost-based pricing, or when the price is set through the profit maximization problem, we have profit maximization pricing. In particular for our survey, the price is set or optimized using game models, auction mechanisms, and the Network Utility Maximization (NUM) problem. Thus there are three pricing models correspondingly, i.e., game-theoretic pricing, auction-based pricing, and Network Utility Maximization (NUM)-based pricing, as shown in Fig.~\ref{taxonomy_pricing_model}.

\subsection{Game-Theoretic Pricing}
\label{subsec:Game_theory}
In game theory, a multi-participant decision scenario is modeled as a game in which each participant as a player selects actions to achieve the best possible payoff while anticipating actions from other players \cite{roy2010survey}. In the following, we briefly present game theoretic-pricing models which have been widely used to improve security in wireless networks.

\subsubsection{Non-cooperative game}
\label{subsec:Game_theory_Non_Cooperative}
In a non-cooperative game, players are selfish. This means that the players maximize only their own payoffs without forming coalitions or making agreements with each other \cite{alskaif2015game}.

Consider a scenario of wireless security involving a legitimate source-destination pair, an eavesdropper, and multiple friendly jammers. The friendly jammers act as the sellers which compete for selling the FJ powers to the source to improve its secrecy capacity. While attempting to maximize their own utilities, the sellers as the players are selfish, and the interactions among them can be modeled as a non-cooperative game. The game is generally defined as a triplet $(\mathcal{N},\mathcal{P},\Pi)$, where $\mathcal{N}$ is the set of $N$ players, $\mathcal{P}$ is the set of strategies of players, and $\Pi$ is the set of payoff functions of the players. Let $p_i$ denote the strategy of player $i$ and $\mathcal{P}_i$ denote the set of strategies of player $i$, then $\mathcal{P}=\mathcal{P}_1 \times \dots \times \mathcal{P}_N$ is namely the Cartesian product of the individual strategy sets. The vector of strategies of $N$ players is $\mathbf{p}=(p_1,\dots,p_N)$, and the vector of corresponding payoffs is $ \boldsymbol{\pi}=(\pi_1(\mathbf{p}),\dots,\pi_N(\mathbf{p})) \in R^N$, where $\pi_i(\mathbf{p})$ is the payoff of player $i$ given the player's chosen strategy and the others' strategies.

\textbf{Definition 1.} A set of strategies or a strategy profile $\mathbf{p^*}=(p_1^*,\dots,p_N^*)$ is called a \textit{Nash equilibrium} if for every player $i$ \cite{friedman1971non},
\begin{equation}
\label{Nash_equilibrium}
 \pi_i(p_i^*,\mathbf{ \overline{p}_{-i}^*}) \geq \pi_i(p_i, \mathbf{\overline{p}_{-i}^*}), \forall p_i \in \mathcal{P}_i,
\end{equation}
where $\mathbf{\overline{p}_{-i}}=(p_1,\dots,p_{i-1}, p_{i+1},\dots, p_N)$ is a vector of strategies of all players except player $i$.

The inequality in (\ref{Nash_equilibrium}) shows that the Nash equilibrium is a stable strategy profile, meaning that none of the players can improve its payoff by unilateral deviation. The reason is that at the Nash equilibrium, all players simultaneously choose the best responses to each other's strategies. In practice, the Nash equilibrium may not always exist, or there may exist multiple Nash equilibria which can make players unclear about which one to choose. Therefore, checking and proving the existence and uniqueness of the Nash equilibrium are important when setting prices based on the non-cooperative game. The existence and uniqueness of the Nash equilibrium can be proved by checking if the strategy of each player is non-empty, compact, and convex, and if its payoff function is a concave or quasi-concave function. Other methods are the Kakutani's fixed point theorem \cite{glicksberg1952further} and the Brouwer's fixed point theorem \cite{kakutani1941generalization}. Another popular method is to prove that the game is a potential game or supermodular game. Because the non-cooperative game models the conflict of players, the attack-defense game was formulated as the non-cooperative game \cite{agah2004non}. The corresponding Nash equilibrium allows intrusion detection systems to have an optimal defense strategy. In the context of wireless security, the non-cooperative game was used to discourage compromised users to launch DoS attacks to legitimate users \cite{shen2016resource}.   

\subsubsection{Stackelberg game}
\label{subsec:Game_theory_Stackelberg}

Different from the non-cooperative game, the Stackelberg game \cite{amir1999stackelberg} is a sequential game in which the players can be leaders or followers. The leaders will make the first move, and then the followers will make corresponding moves based on the leaders' moves. The aim of Stackelberg games is to model sequential multi-agent decision making processes and maximize the profit of the leaders and that of the followers given the strategies of the leaders.

Consider again the scenario in Section \ref{subsec:Game_theory_Non_Cooperative} with two friendly jammers as players, i.e., the sellers. Let $\mathcal{P}_1$ and $\mathcal{P}_2$ denote the sets of pricing strategies of players 1 and 2, respectively. Player 1 selects its pricing strategy $p_1 \in \mathcal{P}_1$ to maximize its payoff function $\pi_1(p_1,p_2)$, and player 2 chooses its pricing strategy $p_2 \in \mathcal{P}_2$ to maximize its payoff function $\pi_2(p_1,p_2)$. Assume that player 2 decides its strategy before player 1 does, then, player 2 is called the \textit{leader}, and player 1 is called the \textit{follower}. We have the following definitions \cite{leitmann2013multicriteria}:

\textbf{Definition 2.} If there exists a function $f: \mathcal{P}_2 \rightarrow \mathcal{P}_1 $ such that, for any fixed $p_2 \in \mathcal{P}_2 $, $\pi_1(fp_2,p_2) \geq \pi_1(p_1,p_2) $, $\forall p_1 \in \mathcal{P}_1 $, where $fp_2$ is the best response of player 1 given $p_2$, and if there exists $p_{2s2} \in \mathcal{P}_2 $ such that $\pi_2(fp_{2s2},p_{2s2}) \geq \pi_2(fp_2,p_2) $, then the pair $(p_{1s2}, p_{2s2}) \in \mathcal{P}_1 \times \mathcal{P}_2$, where $p_{1s2}= fp_{2s2} $, is called a \textit{Stackelberg strategy pair}.

In other words, the Stackelberg strategy is optimal for the leader when the follower responds to the leader with the optimal strategy. As presented in \cite{nguyen2017resource}, the payoff of the leader in the Stackelberg solution is guaranteed to be no less than that in the Nash solution. The reason is that by choosing the Stackelberg strategy, the leader imposes a favorable solution to itself. This feature makes the Stackelberg game be an efficient solution for the network security. For example, it allows a defender to decide an optimal strategy after observing the strategy of an attacker \cite{chen2009game}. Also, the game optimizes the defender's utility if the attacker deviates from equilibrium strategies \cite{an2011refinement}. 

\subsection{Auction-Based Pricing}
\label{subsec:Game_theory_auction}
An auction can be used as a pricing mechanism for selling commodities, i.e., resources in wireless security context, and setting corresponding prices according to specific rules. Definitions of common terminologies in auctions such as bidder, seller, auctioneer, ask, and bid can be found in \cite{luong2016data}. Also, a tutorial of auctions used for resource management in wireless networks was provided in \cite{zhang2013auction}. In what follows, we discuss the auctions which have been used to improve the wireless security. 

\subsubsection{English and Dutch auctions}
\label{subsec:Game_theory_auction_English_Dutch}
English and Dutch auctions are known as the \textit{open-outcry} auctions in which bids of buyers are disclosed to each other during the auction~\cite{vijay2002auction}.
\begin{itemize}
\item \textit{English auction:} The English auction is an ascending-bid auction performed in several rounds. The auctioneer initializes the lowest price of the resource and then increases the price in the next rounds. The auctioneer terminates the auction if there is no new higher bidding price submitted by any buyer. The buyer with the highest price wins the resource and pays the price not less than the lowest price that the seller can accept. The highest price is also called a \textit{hammer price}. 
\item \textit{Dutch auction:} The Dutch auction is a descending-bid auction in which the auctioneer or seller initializes a high price for the resource and then decreases the price until one buyer accepts the price. 
\end{itemize}

Compared to the sealed-bid auction discussed in the next section, the English and Dutch auctions have been used less frequently for the wireless security due to the disclosed bids. However, with their simplicity, they were still adopted to quickly revoke the malicious users as proposed in \cite{reidt2009fable}. 

\subsubsection{Sealed-bid auction}
\label{subsec:Game_theory_auction_sealed_bid}
Different from the English and Dutch auctions, bidders in a sealed-bid auction submit sealed bids to the auctioneer such that no bidder knows the bid information of the others. The first- and second-price sealed-bid auctions are two most common types of the sealed-bid auction. 

\begin{itemize}
\item \textit{First-price sealed-bid auction:} The bidder with the highest price wins the commodity, and the winner pays the seller the highest price. 
\item \textit{Second-price sealed-bid auction or Vickrey auction:} In this auction, the winner only pays the second-highest price rather than the highest price that it submitted \cite{lucking2000vickrey}. Since the winner pays the price less than its expected price, the Vickrey auction motivates buyers to bid its true valuation on the resource. Such an auction is thus called a \textit{strategy-proof}, \textit{truthful}, or \textit{ incentive compatible} mechanism. This feature enables the Vickrey auction to be widely used for the wireless security to prevent the misbehaviors of users.
 
\item \textit{Vickrey-Clarke-Groves (VCG) auction:} The VCG auction is the extension of the Vickrey auction with multiple commodities. In the VCG auction, the commodities are allocated in a socially optimal manner, and the price that the winner pays the seller equals the loss of the social value caused by its winning the commodity. Since each winner is charged by only the marginal harm to other bidders, the VCG auction enables bidders to submit true valuations in their bids. The VCG auction is thus strategy-proof. However, since its objective is to maximize the social welfare, the price set by the VCG auction is sometimes too low. Thus the losers in the auction can afford the prices. For example, the losers can collude to win channels and then sublease the channels to others. This concern can be resolved by integrating the VCG auction with encryption methods as proposed in \cite{pan2012using}. 
\end{itemize}
\subsubsection{Double auction}
\label{subsec:Game_theory_auction_double}
In a double auction, buyers and sellers simultaneously submit their bids, i.e., bidding prices, and asks, i.e., asking prices, to an auctioneer, respectively \cite{friedman1993double}. Upon receiving the bids and asks, the auctioneer performs matching between the sellers' asks and the buyers' bids as follows. It first sorts the buyers' bids in a descending order and the sellers' asks in an ascending order. The auctioneer then finds the largest index $m$ at which the ask $p^a_m$ is less than or equal to the bid $p^b_m$, i.e., $p^a_m \leq p^b_m$. The transaction price $p^*=(p^a_m + p^b_m)/2$ is called a \textit{clearing price}. The corresponding buyer receives the resource, and the seller receives the payment $p^*$. The process is repeated to match the remaining buyers and sellers and to determine corresponding transaction prices. 

Compared to the VCG auction, the double auction achieves more desirable properties. In addition to the \textit{truthfulness} or \textit{incentive compatibility}, the double auction holds other properties such as the \textit{individual rationality}, i.e., no participant loses when joining the auction, and \textit{balanced budget}, i.e., the auctioneer gains positive benefit. Because of these properties, the double auction was used to achieve the $k$-anonymity location privacy as proposed in \cite{zhang2016designing}. 

\subsubsection{Share auction}
\label{subsec:Game_theory_auction_share}
Share auction is a market mechanism for allocating a perfectly divisible resource, e.g., the power, among bidders \cite{huang2010game}. In the share auction, bidders submit their bids, i.e., amount of requested resource, to the auctioneer. Then, the resource allocation and the price for each bidder are proportional to their bids. Consider a specific model involving $N$ source-destination pairs, a friendly jammer and an eavesdropper. Sources as buyers, i.e., bidders, submit their bids to compete for the FJ power $P_i$ from the friendly jammer, i.e., the seller, to increase their secrecy capacities. Based on the share auction, the friendly jammer allocates its power to source $i$ as follows:
\begin{equation} 
\label{share_auction}
P_i=\frac{b_iP_{\text{max}}}{\beta + \sum_{j=1}^{N}b_j},
\end{equation}
where $P_{\text{max}}$ is the maximum power of the friendly jammer, $b_i$ is the bid of source $i$, and $\beta$ is any positive constant. Source $i$ then pays the jammer a price $p_i=\lambda P_i$, where $\lambda$ is the price per unit of power. To determine the optimal bid $b_i^*$, source $i$ finds $b_i$ so as to maximize its utility $U_i$. $U_i$ is the difference between the secrecy capacity change due to power $P_i$ and price $p_i$. If $\lambda$ is set to an appropriate value, then $U_i$ is quasi-concave within a feasible region, and there exists an optimal bid $b_i^*$ to optimize the source's utility \cite{zhu2009improved}. The desirable outcome of the auction is the Nash equilibrium such that no source has an incentive to deviate its bidding strategy unilaterally. 
\subsubsection{Ascending Clock Auction (ACA)}
\label{subsec:Game_theory_ACA}
ACA is a type of multiple-round auctions. In each round, the auctioneer announces a price, and bidders submit their demands at the given price. The
auctioneer increases the price until the total demand meets the supply. To further understand the ACA process, we consider a model with multiple sources as bidders, i.e., buyers, a friendly jammer as a seller, and an eavesdropper. The FJ power $P_{\text{max}}$ is traded as a single object, and the sources only bid zero or $P_{\text{max}}$. First, the friendly jammer sets an initial asking price $\lambda^0$ and broadcasts this price to all the sources. Source $i$ calculates the maximum utility, i.e., $U_i^*$, that it can obtain from purchasing $P_{\text{max}}$. If $U_i^* \leq0$, source $i$ will bid zero power. Otherwise, it will bid $P_{\text{max}}$. If there are more than one source bidding with $P_{\text{max}}$, the friendly jammer increases the asking price and then announces this price in the next auction round. Given the new price, each source recalculates its maximal utility to decide if it bids zero or $P_{\text{max}}$. This process is repeated until there is only one source bidding $P_{\text{max}}$. Generally, with the ACA, the bidders reveal their demands but not prices to the auctioneer. Thus in addition to improving the secrecy capacity as discussed in this section, the ACA was used to preserve the privacy for bidders as proposed in \cite{rathinakumar2016gavel}. 


The summary of the above auctions along with their applications
for the wireless security is given in Table~\ref{table_auction_sum}. As seen,
auction mechanisms have been adopted to address various security issues in wireless networks. Moreover, the Vickrey auction has been more commonly used compared with the other auctions due to its truthfulness guarantee. 

\begin{table*}[h]
\caption{A summary of key features and suitable scenarios of auction mechanisms used for security issues in wireless networks}
\label{table_auction_sum}
\scriptsize

\begin{centering}
\begin{tabular}{|>{\centering\arraybackslash}m{2.2cm}|>{\centering\arraybackslash}m{3.3cm}|>{\centering\arraybackslash}m{6.1cm}|>{\centering\arraybackslash}m{3.3cm}|>{\centering\arraybackslash}m{1.3cm}|}
\hline
\cellcolor{mygray} &\cellcolor{mygray} &\cellcolor{mygray} &\cellcolor{mygray} &\cellcolor{mygray} \tabularnewline
\cellcolor{mygray} \multirow{-2 }{*}{\textbf{Auction type}} &\cellcolor{mygray} \multirow{-2}{*} {\textbf{Market structure}} &\cellcolor{mygray} \multirow{-2}{*} {\textbf{Key descriptions}} &\cellcolor{mygray} \multirow{-2}{*} {\textbf{Suitable scenarios}} &\cellcolor{mygray} \multirow{-2}{*}{\textbf{Solution}} \tabularnewline
\hline
\hline
 English auction \cite{vijay2002auction} & \begin{itemize} \item {\textit{Single-sided auction:} A seller and multiple buyers}  \end{itemize} &  \begin{itemize} \item{Open-outcry ascending-price auction: The winner pays the second highest price} \end{itemize}&\begin{itemize} \item {Information integrity} \end{itemize} &Nash equilibrium\tabularnewline \cline{2-5}
\hline
Dutch auction \cite{vijay2002auction} & \begin{itemize} \item{\textit{Single-sided auction:} A seller and multiple buyers} \end{itemize} & \begin{itemize} \item{Open-outcry descending price auction: The winner pays the final price} \end{itemize} &\begin{itemize}  \item{Black hole attack} \end{itemize} &Nash equilibrium\tabularnewline \cline{2-5}
\hline
First-price sealed-bid auction \cite{lucking2000vickrey}& \begin{itemize} \item{\textit{Single-sided auction:} A seller and multiple buyers} \end{itemize} & \begin{itemize} \item{Sealed-bid auction: The winner pays the highest price}\end{itemize} &\begin{itemize} \item {Black hole attack} \item{Privacy concerns}  \item{Faked sensing attack} \end{itemize} &Nash equilibrium\tabularnewline \cline{2-5}
\hline
Vickrey auction \cite{lucking2000vickrey}& \begin{itemize} \item{\textit{Single-sided auction:} A seller and multiple buyers} \end{itemize} & \begin{itemize} \item{Sealed-bid auction: The winner pays the second highest price}\end{itemize} &\begin{itemize} \item {Eavesdropping attack} \item{Privacy concerns}  \item{User collusion/bid-rigging} \end{itemize} &Nash equilibrium\tabularnewline \cline{2-5}
\hline
Vickrey-Clarke-Groves (VCG) auction \cite{ausubel2006lovely}& \begin{itemize} \item{\textit{Single-sided auction:} A seller and multiple buyers} \end{itemize}& \begin{itemize} \item{Generation of the Vickrey auction with multiple commodities: The winner pays the price equal to the loss of the social welfare due to its getting commodities} \end{itemize} &\begin{itemize} \item {Eavesdropping attack}  \item{User collusion/bid-rigging} \end{itemize} &Bayesian Nash
equilibrium\tabularnewline \cline{2-5}
\hline
Share auction \cite{huang2007auction}& \begin{itemize} \item{\textit{Single-sided auction:} A seller and multiple buyers} \end{itemize} & \begin{itemize} \item{Resources are allocated to buyers according to the ratio of the buyers' bids} \end{itemize} &\begin{itemize} \item {Eavesdropping attack}  \item {DDoS attack} \end{itemize} &Nash equilibrium\tabularnewline \cline{2-5}
\hline
Ascending Clock Auction (ACA) \cite{zhang2013ascending}& \begin{itemize} \item{\textit{Single-sided auction:} A seller and multiple buyers} \end{itemize} & \begin{itemize} \item{The resource price is increased at each round until the total demand equals the total supply} \end{itemize} &\begin{itemize} \item {Eavesdropping attack} \item{False-name
bids} \end{itemize} &Walrasian equilibrium\tabularnewline \cline{2-5}
\hline
 Double auction\cite{friedman1993double}& \begin{itemize} \item{\textit{Double-sided auction:} Multiple sellers and multiple buyers} \end{itemize}& \begin{itemize} \item{The auctioneer matches sellers' asks and buyers' bids} \end{itemize} &\begin{itemize} \item {Eavesdropping attack} \item{Privacy concerns} \end{itemize} &Market
equilibrium\tabularnewline \cline{2-5}
\hline
\end{tabular}
\par\end{centering}
\end{table*}

\subsection{Network Utility Maximization (NUM)-based Pricing}
\label{subsec:NUM_problem}

NUM is known as a dual-based distributed algorithm for the network resource allocation which aims at maximizing the social welfare subject to the resource constraint of the network \cite{mas1995microeconomic}. In the standard NUM problem, the social welfare is the sum of utility functions of all users in the network. In the context of wireless security, the utility functions can be the secrecy capacities of sources. When the sources employ the FJ powers from the friendly jammer to improve their secrecy capacities, a certain cost is incurred to the friendly jammer. Thus the total cost needs to be introduced in the standard NUM problem, and the new problem is namely \textit{modified NUM problem}. As discussed in \cite{nguyen2017resource}, to efficiently solve the modified NUM problem, the pricing-based iterative solutions can be adopted. Specifically, consider a model with $N$ sources as buyers which reserve powers from a friendly jammer, i.e., a seller, to improve their secrecy capacities. Assume that the friendly jammer charges source $i$ a price per unit of power $\lambda_i$. Given the price vector $\boldsymbol{\lambda}=(\lambda_1,\dots,\lambda_N)$, the modified NUM problem is given by
\begin{equation}
\label{NUM_SYS_PRICE}
\max \limits_{\textbf{P}, \sum_{i}^{N}P_i \leq P_{\text{max}}} \sum_{i}^{N}(\Delta C_i(P_i)-\lambda_iP_i) + (\boldsymbol{\lambda}^\top \textbf{P}-c(\textbf{P})),
\end{equation}
where $P_i$ is the power that source $i$ receives, $\textbf{P}=(P_1,\dots,P_N)$, $\Delta C_i(P_i)$ is source $i$'s secrecy capacity change due to $P_i$, $c(\textbf{P})$ is the total cost which is incurred to the friendly jammer when it applies power $\textbf{P}$, $U_i=(\Delta C_i(P_i)-\lambda_iP_i)$ is actually the utility function of source $i$, and $(\boldsymbol{\lambda}^\top \textbf{P}-c(\textbf{P}))$ is the profit of the friendly jammer. Source $i$ determines $P_i$ by solving the following problem
\begin{equation}
\label{NUM_SYS_PRICE_User}
P_i(\lambda_i)= \arg \max \limits_{P_i\in D_i} U_i, i=1,\dots, N,
\end{equation}
where $D_i$ represents a continuous range of power. Generally, $U_i$ may not be a concave function, and (\ref{NUM_SYS_PRICE_User}) may not be the convex optimization problem. However, as stated in Section~\ref{subsec:Game_theory_auction_share}, if the friendly jammer sets the price $\lambda_i$ to the optimal value, the utility function $U_i$ is a quasi-concave shape function within the feasible interval. Thus there exists an optimal $P_i$ to maximize its utility. The optimal $P_i$ can be obtained using iterative methods. 

Apart from improving the secrecy capacity as discussed above, NUM was also adopted as distributed solutions to prevent DDoS attacks as proposed in \cite{chorppathbayesian}. 

\textbf{Summary:} In this section, we present basics of economic and pricing theory used to address wireless security issues. Specifically, we provide definitions, mechanism descriptions, and the rationale behind the use of economic and pricing models for the wireless security. Additionally, we give comparisons between the models. In the next sections, we review economic and pricing approaches for various security issues in wireless networks. 

\section{Eavesdropping attack}
\label{sec:Survey_eavesdropper_attack}
Due to the broadcast nature of the wireless transmission, anyone within the communication
range can intercept the source's information. An unauthorized receiver is called an eavesdropper \cite{hao2013multihop}. Power allocation is the major approach used to improve secrecy capacity in that a friendly jammer allocates the Friendly Jamming (FJ) power to source-destination links. This section reviews the applications of economic and pricing models to maximize the secrecy capacity in wireless networks. Note that the eavesdropper mentioned in this section is passive. More specifically, two types of eavesdroppers are discussed in this section: 
\begin{itemize}
\item{\textit{External eavesdropper}:} An external eavesdropper is an unauthorized receiver which is not an intermediate node in any path which is selected for relaying or forwarding the information \cite{hao2013multihop}. Centralized schemes in which the objective is to optimize the secrecy capacity can be applied. However, they suffer from the fact that the friendly jammer needs to know exactly all the private information of the sources, destinations, and eavesdropper. Economic and pricing models as distributed solutions can maximize the secrecy capacity without requiring the perfect knowledge of all channel information. 

\item{\textit{Internal eavesdropper}:} An internal eavesdropper is an intermediate node in a path which is selected for relaying or forwarding the information \cite{hao2013multihop}. It may be an untrusted relay node which tries to eavesdrop the information coming from the source. Economic and pricing models provide incentive mechanisms to friendly jammers to contribute power to interfere the internal eavesdropper. 
 \end{itemize}

\subsection{External Eavesdropper}
\label{sec:Survey_eavesdropper_attack_external}
This section reviews the applications of economic and pricing models for the FJ power allocation between the friendly jammer and the source-destination pairs to maximize the secrecy capacity. Economic and pricing models are developed for different senarios. For example, when the model involves only one friendly jammer and multiple source-destination pairs, the single-side auction schemes can be used. For multiple friendly jammers and multiple source-destination pairs, the double auction or matching theory can be adopted. 

\subsubsection{Share auction}
\label{sec:Survey_eavesdropper_attack_external_SNR_auction}
\begin{figure}[t!]
 \centering
\includegraphics[width=5.5cm, height=6cm]{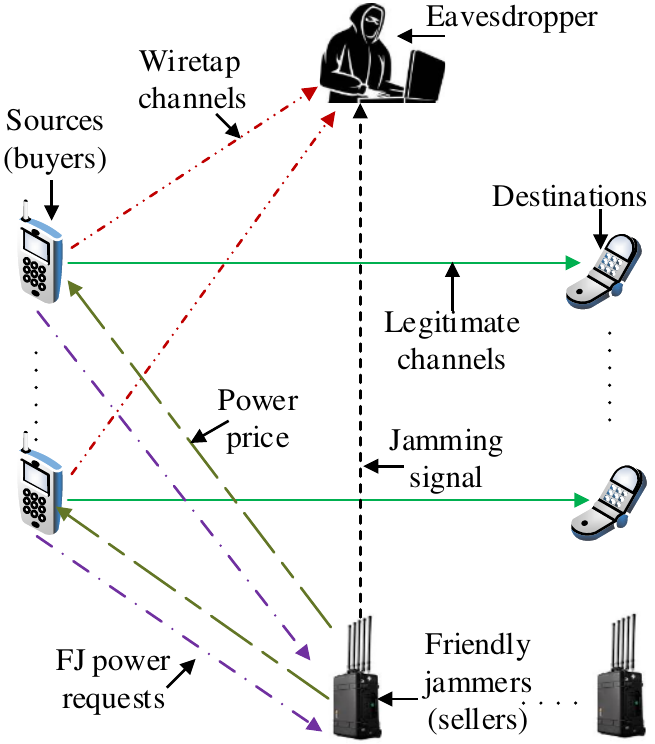}
 \caption{FJ power allocation for multiple source-destination pairs based on pricing models.}
 \label{eavesdropper_SNR_auction}
\end{figure}
The authors in \cite{zhu2009improved} applied the share auction for the FJ power allocation to optimize the secrecy capacity for sources. In the share auction, the FJ power is split among sources, and their payments depend solely on the bids \cite{huang2010game}. The share auction is an appropriate solution since it allows to allocate the FJ power to the sources in a distributed manner with the limited local information. The model is shown in Fig.~\ref{eavesdropper_SNR_auction} which consists of multiple source-destination pairs, one friendly jammer, and one eavesdropper. The sources as bidders compete for the power from the friendly jammer as the seller. First, each source determines an amount of power as a bid to maximize its utility, which is the difference between the secrecy capacity change and the price that the source pays the firendly jammer. The source's secrecy capacity is defined as in Section~\ref{sec:Intro_Security_Physical_layer}. Given the sources' bids, the friendly jammer allocates its power and charges the sources according to the allocated power. The simulation results showed that the secrecy capacity for the source obtained by the proposed solution is close to that of the centralized approaches. However, the proposed solution does not require the exact knowledge of channel information which is difficult to achieve in practice. 

\subsubsection{Traditional Ascending Clock Auction (ACA-T) and Alternative Ascending Clock Auction (ACA-A)}
\label{sec:Survey_eavesdropper_attack_external_ACA_T_ACA_A}
Different distributed auctions can be found in \cite{zhang2013ascending} for the FJ power allocation, that are ACA-T and ACA-A. They are based on the standard ACA as presented in Section \ref{subsec:Game_theory_auction}. The auctions are used for the friendly jammer as the seller, i.e., the auctioneer, to allocate the power to sources as bidders. 

In ACA-T, the friendly jammer initially announces a reserve price. Sources determine power demands to optimize their utilities, each of which is defined similar to that in \cite{zhu2009improved}. The sources then submit their demands to the friendly jammer. The friendly jammer iteratively adjusts the price so that the total demand of the sources equals the maximum power.  To ensure that the total demand equals the friendly jammer's maximum power, i.e., the full utilization of the power, the final power for each source is allocated according to the proportional-rationing rule \cite{ausubel2004efficient}. 

The ACA-T scheme guarantees the efficiency of the power allocation. However, the sources are proved to have an incentive to misreport their true demands since this can lead to a greater utility. Thus the ACA-T scheme is not cheat-proof. To address this problem, the ACA-A solution \cite{ausubel2004efficient} was adopted. Generally, the ACA-A has the same procedures as the ACA-T. However, at every iteration in the ACA-A scheme, the friendly jammer calculates the cumulative clinch \cite{ausubel2004efficient}, which is the amount of power that each source is guaranteed to win at the current iteration. The payment from each source is then determined based on its cumulative clinch. It was proved that the best strategy of the source at every iteration is to report its true optimal demand given other sources' true demands. As shown in the simulation, the secrecy rates of both ACA-A and ACA-T are much greater than that of the no-jamming case. Also, the source's maximum utility in the ACA-A is achieved at its true optimal demand while that in the ACA-T is at a value less than its true optimal demand. In other words, the ACA-A is cheat-proof while the ACA-T is not. 

Given the advantages of the ACA-A scheme, the authors in \cite{li2015secure} employed this auction for the joint subcarrier and FJ power allocation to improve the uplink secrecy rate in cellular networks. The model consists of one base station, multiple mobile users, and one eavesdropper. The base station as an auctioneer allocates its subcarriers and FJ power to the mobile users as bidders. The auction process is similar to that in \cite{zhang2013ascending}. However, subcarrier prices are introduced in addition to the power price. Moreover, the Lyapunov's theorem \cite{shakkottai2008network} was adopted to prove the convergence of the proposed scheme to the optimal solutions of the subcarrier, FJ power, and prices. 

\subsubsection{Stackelberg game}
\label{sec:Survey_eavesdropper_attack_external_Stackelberg}
To maximize the utilities for both the sources and the friendly jammer, the authors in \cite{wang2013stackelberg} adopted the Stackelberg game (see Section~\ref{subsec:Game_theory_Stackelberg}). The model is similar to that in \cite{zhu2009improved} where the sources act as followers, i.e., buyers, and the friendly jammer is the leader, i.e., the seller. Given the power price of the friendly jammer, each source determines the power demand so as to maximize its utility. The source's utility is defined similar to that in \cite{zhu2009improved}, which was proved to be concave in the power. Then, the optimal power demand is calculated by taking the first-order derivative of the utility function. Given the sources' demands, the friendly jammer determines the power prices to maximize its utility, i.e., the total payment from all sources, using the first-order derivative. It was indicated that the best-response function of the prices is a standard function, meaning that it satisfies the conditions including positivity, monotonicity, and scalability. Therefore, there exists a unique Stackelberg equilibrium which is the pair of the optimal FJ power and power price. The numerical results showed that the sum-secrecy rate of the proposed approach is significantly improved compared with that of the equal-power allocation algorithm. However, the utility improvement of the friendly jammer was not demonstrated. 

The same approach can be found in \cite{yue2012fairness}. However, in addition to the secrecy rate, the authors in \cite{yue2012fairness} considered guaranteeing the fairness among sources and balancing their utilities. Accordingly, in the first stage of the game, the sources determine their FJ power demands so as to balance obtained utilities among them through guaranteeing the Kalai-Smorodinsky Bargaining Solution (KSBS) \cite{kalai1975other}. The power for each source is then calculated based on the property of the KSBS and the power constraints. The simulation results illustrated that the proposed scheme has the similar fairness but much higher total utility of the sources than that of the equal-power allocation algorithm. 

To further enhance the utility for one source, multiple friendly jammers are employed as presented in \cite{han2009physical}. Unlike \cite{yue2012fairness}, the source in this model acts as a leader, i.e., the buyer, while friendly jammers are followers, i.e., sellers. The source first determines optimal amounts of power from the friendly jammers to maximize its utility, and then the friendly jammers set the optimal prices to maximize their payments. The simulation results showed that the utility of the source is close to that of the centralized approach regardless of friendly jammers' locations. One of the reasons is that the source can switch among friendly jammers for its best performance.

The Stackelberg game can be adopted to improve the secrecy capacity in relay networks as proposed in \cite{maorui2011security}. The model involves one source-destination pair, one eavesdropper, and one relay node. Since the relay node can communicate with both the source and the destination, it can perform friendly jamming. The relay node is thus called \textit{friendly relay}. The relay acts as the follower, i.e., the seller, which provisions the information relay and friendly jamming for the source as the leader, i.e., the buyer. The optimal power for the source and the optimal price for the relay were then determined in a similar way as in \cite{han2009physical}. 

\subsubsection{Vicrkey auction}
\label{sec:Survey_eavesdropper_attack_external_Vickrey_auction}
To maximize the secrecy capacity and to guarantee the truthfulness of the FJ power allocation, the Vickrey auction can be used. 

\begin{figure}[t!]
 \centering
\includegraphics[width=7.4cm, height=7.3cm]{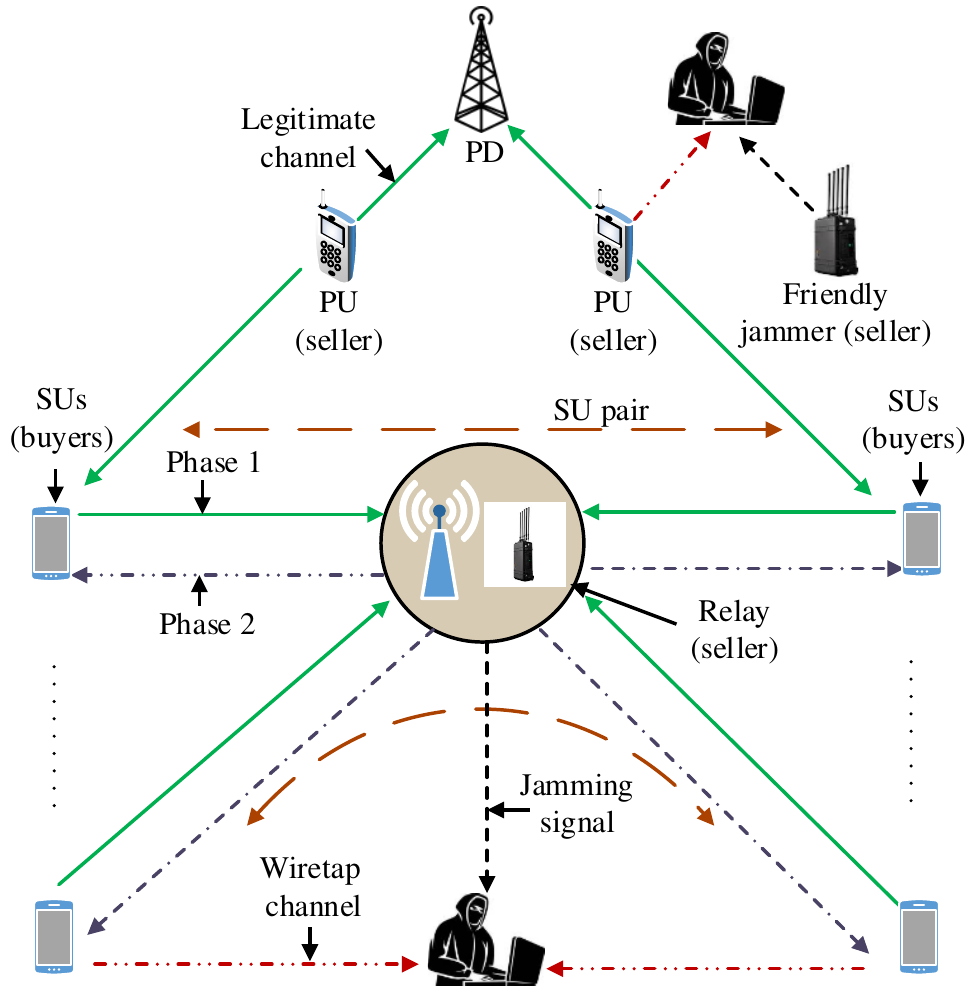}
 \caption{FJ power allocation in cognitive two-way relay networks.}
 \label{external_Eavesdropper_Vickrey}
\end{figure}

The first Vickrey auction approach was proposed in \cite{wang2012power} to improve the secrecy capacity for the communication of Secondary User (SU) pairs in a cognitive two-way relay network. The model involves one Primary User (PU), multiple SU pairs, one relay, and one eavesdropper as shown in Fig.~\ref{external_Eavesdropper_Vickrey}. The relay acts as an auctioneer (seller) which provides the relay and FJ powers to the sources of SU pairs, i.e., bidders (buyers). The total power is divided into $K$ equal units. Each source first calculates its marginal secrecy rate when obtaining $k^i \leq K$ power units. The source then submits its bid which includes the vector of $K$ marginal secrecy rates and the number of desired power units to the relay. The $K$ highest bids are deemed winning and the corresponding sources obtain desired power units. Source $i$ which wins $k^i$ units pays the $k^i$ highest losing bids of the other sources. However, how each source determines its desired power units was not explained.

The allocation in \cite{wang2012power} may be less efficient if the total power demand of winners is less than the maximum power of the relay due to the unused power units. Therefore, the authors in \cite{wang2011improve} adopted the sequential Vickrey auction which allocates sequentially the power units to the sources until no power unit remains. The sequential Vickrey auction repeats several rounds, and each round is a typical Vickrey auction. The proposed scheme improves the system efficiency and the total revenue for the relay compared with \cite{wang2012power}. However, the proposed scheme is recursive, and its complexity is high when the number of power units is large. 
 
Due to the resource constraint of one relay, multiple relays may be used. The problem is to select $K$ relays to maximize the total secrecy rate for the source. The authors in \cite{deng2013truthful} applied the VCG auction, i.e., a generalization of the Vickrey auction for multiple items, to the model with one source-destination pair, one eavesdropper, and multiple relays. First, relays as sellers report the channel information to the source as the buyer. The source calculates the secrecy rate for each relay and selects $K$ relays which provide the maximum sum of secrecy rates. The selected relay receives the total payoff including: (i) the expected payoff and (ii) the \textit{transfer payment}. The expected payoff is proportional to the price per unit of secrecy rate achieved by the relay and the probability of the relay being chosen. The transfer payment of the relay is the difference in the total expected payoff of other relays with and without the relay in the system. The payment strategy enables the proposed approach to achieve the incentive compatibility. The simulation results showed that when a relay's real secrecy rate increases, i.e., higher probability of true report, its total payoff increases.

The VCG-based scheme in \cite{deng2013truthful} cannot achieve the balanced budget since the sum of all selected relays' transfer payments is less than zero. To address this issue, the authors in \cite{liu2011enforce} proposed a new transfer payment based on the sum of other relays' expected payoffs given the reported information of one relay. It was proved that the new transfer payment can perform a payment reallocation among relays, and the total transfer payment of all relays is zero. Therefore, the system can achieve the balanced budget. 

Using the model in \cite{han2009physical}, the authors in \cite{stanojev2011cooperative} applies the Vickrey auction for the joint bandwidth and FJ power allocation. Generally, the source selects a friendly jammer which maximizes the source's secrecy rate and then allocates the access bandwidth to the friendly jammer. Specifically, the source as the seller announces an amount of bandwidth that can be allocated to a friendly jammer. The friendly jammers as the bidders determine their optimal FJ powers, i.e., bids. Given the bids, the source calculates the corresponding secrecy rates and selects a friendly jammer with the highest secrecy rate as the winner. The winner pays the source the FJ power so that the corresponding secrecy rate is at least the second-largest one. The simulation result showed that the proposed solution outperforms the approach based on the Stackelberg game \cite{osborne1994course} in terms of the average secrecy rate. 

A joint time slot and FJ power allocation can be found in \cite{ma2016improving}. This approach aims to improve the secrecy rate for the PU-Primary Destination (PD) pair in Cognitive Raio Networks (CRNs). The model consists of one PU-PD pair, multiple SUs, and one eavesdropper. The PU acts as an auctioneer, i.e., a seller, which selects one of SUs, i.e., bidders, for the jamming service. Accordingly, each SU submits its bid including (i) the FJ power, (ii) a time fraction, i.e., the time for the SU to transmit the PU's data to the PD, and (iii) the channel information related to the SU. The PU calculates the secrecy rate corresponding to each SU. The SU which can provide the highest secrecy rate is selected as the winner. It then provides the PU the second-highest bid, and receives a time slot as a reward. In fact, to enhance the efficiency of the time slot allocation, the PU should incorporate time slot requests of SUs when evaluating bids, which is not done in \cite{ma2016improving}. 


\subsubsection{Matching theory}
\label{sec:Survey_eavesdropper_attack_external_matching}
Consider a more general scenario including multiple source-destination pairs, multiple friendly jammers, and one eavesdropper. The authors in \cite{bayat2012distributed} adopted the matching theory \cite{sotomayor1990two} to match each source with each friendly jammer to maximize the sum of utilities of sources and friendly jammers. First, friendly jammers as sellers announce their FJ power and offered prices. Then, each source as the buyer calculates its utility with each friendly jammer. The utility is the source's secrecy rate minus the price that the source pays the friendly jammer. The source bids for its friendly jammer which can maximize the source's utility. If the friendly jammer receives only one bid, the friendly jammer and the corresponding source will match together. If the friendly jammer receives more than one bid, it will increase the offered price. This process is repeated until there is no new offered price from any friendly jammer. The simulation results showed that the proposed scheme improves significantly the average secrecy rate and social welfare compared with the random matching \cite{aliprantis2007random}. However, the complexity of the proposed algorithm is much higher, especially when the number of sources or friendly jammers is large.

\subsubsection{Double auction}
\label{sec:Survey_eavesdropper_attack_external_double}
The authors in \cite{wang2015dasi} extended the model in \cite{ma2016improving} with multiple eavesdroppers and multiple PUs. SUs provision the FJ power to PUs to maximize the secrecy rates for the communications between the PUs and the Base Station (BS) while guaranteeing some economic properties such as truthfulness, individual rationality, and balanced budget. The double auction is thus an appropriate solution. In this model, PUs are sellers, SUs are buyers, and the BS is the auctioneer. A PU's ask, i.e., asking price, is its minimum secrecy rate requirement, and an SU's bid, i.e, bidding price, is a vector of all PUs' secrecy rates that the SU can provide under its minimum bandwidth requirement. After receiving PUs' asks and SUs' bids, the BS matches one PU to at most one potential SU by using the maximum weighted matching algorithm \cite{yang2011opra}. Then, the BS determines the winning PUs and SUs as well as the clearing price similar to those in Section \ref{subsec:Game_theory_auction_double}. The winning PUs receive the winning SUs' power, and the winning SUs receive the winning PUs' access bandwidth. 

The model in \cite{wang2015dasi} was designed for a static CRN. In real environments, the SUs may join and leave the networks dynamically, and thus a dynamic double auction can be used as proposed in \cite{wang2016auction}. Accordingly, the auction period is divided into multiple time slots. Since the bid of an SU includes the arrival and departure time in addition to the bidding price, the auction needs to ensure that the SU cannot obtain higher utility by cheating its arrival or departure time. To address this issue, a \textit{preservation price} was introduced which particularly depends on the reported arrival and departure time of the SU. An SU is selected for participating in the auction if the SU has a bidding price larger than or equal to its preservation price. The user assignment, winner determination, pricing mechanism are then applied similar to those in \cite{wang2015dasi}. However, the proposed solution in \cite{wang2016auction} is only applicable to the scenario where the SUs rent a channel for one time slot. The long-term leasing scenario with multiple time slots needs to be considered.

\subsubsection{Non-cooperative game}
\label{sec:Survey_eavesdropper_attack_external_non_cooperative}
In practice, a source can be equipped with multiple antennas, and it can generate the FJ signal along with the information signal without requiring an external friendly jammer. The authors in \cite{siyariprice} employed the multiple antennas to improve the secrecy rate for two source-destination pairs in the presence of one eavesdropper. To discourage the sources from acting selfishly which may interfere with each other, the non-cooperative game and a pricing factor, i.e., a price per unit of power, were adopted. Accordingly, the strategy of each source is to find its FJ power to maximize the difference between its secrecy rate and the price that it pays. The optimal solution is obtained using the KKT conditions \cite{boyd2004convex}. By iteratively using the pricing factor to set the power for both sources, the game converges to the Nash equilibrium. However, such optimal solutions of the price and FJ power require the perfect knowledge of the eavesdropping channel which is challenging in practice.

\subsection{Internal Eavesdropper}
\label{sec:Survey_eavesdropper_attack_internal}
An internal eavesdropper typically refers to an untrusted relay in relay networks. The following discusses how to apply pricing models for the interaction between sources and friendly jammers to prevent eavesdropping from untrusted relays.  
\subsubsection{Utility maximization}
\label{sec:Survey_eavesdropper_attack_internal_utility_maximization}
\begin{figure}[t!]
 \centering
\includegraphics[width=7.5cm, height=5.7cm]{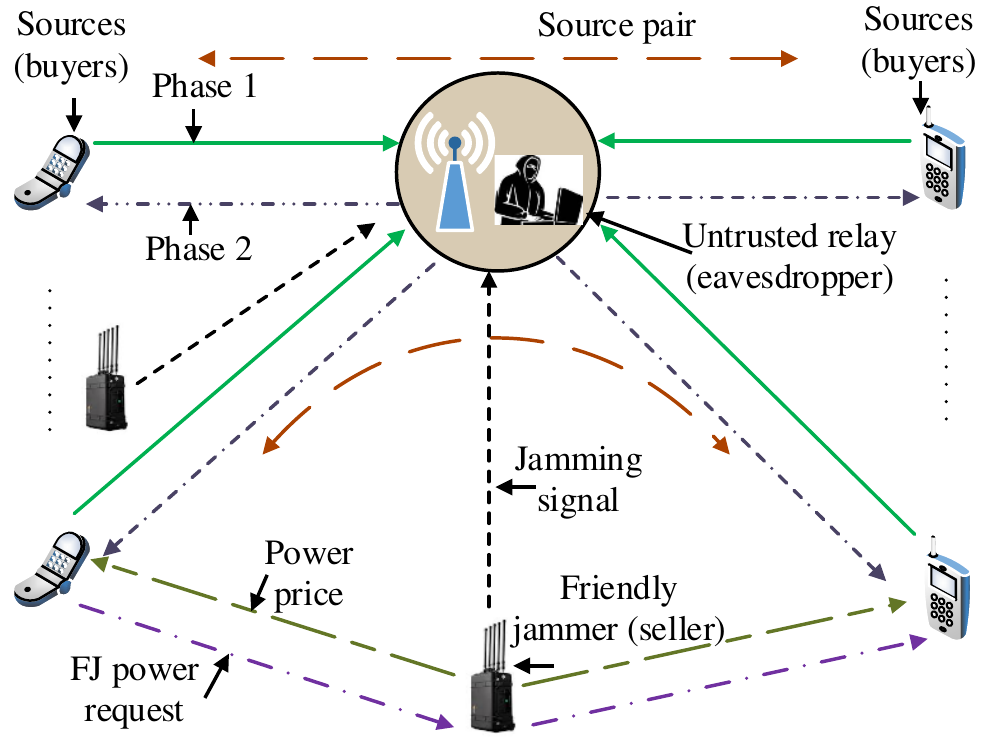}
 \caption{FJ power allocation in a multiuser two-way untrusted relay network.}
 \label{eavesdropper_ACA_A_multi_source_relay}
\end{figure}

The authors in \cite{zhang2010physical} aim to maximize the secrecy rate for the two-way untrusted relay network using the utility maximization problem. Fig.~\ref{eavesdropper_ACA_A_multi_source_relay} shows such a general network with multiple pairs of two sources. The considered model consists of one untrusted relay, i.e, the eavesdropper, one pair of two sources, and multiple friendly jammers. The sources as buyers buy power from the friendly jammers, i.e., sellers, to maximize the sum of sources' utilities. The sum of sources' utilities can be considered to be the source pair's utility which is the sum of secrecy rates of the two sources minus the total price that the sources pay the friendly jammer. The optimal power of each friendly jammer is then determined using the first-order derivative. The friendly jammer's optimal power depends on its price and the other friendly jammers' power. As shown in the simulation results, when the price is high, the bought power reduces or even becomes close to zero. However, if the price is low, the friendly jammers receive low benefit. Thus cooperative games can be applied to achieve the optimal solutions of FJ power and price.

\subsubsection{Stackelberg game}
\label{sec:Survey_eavesdropper_attack_internal_utility_maximization}
To improve utilities of sources and friendly jammers in the two-way untrusted relay system, the Stackelberg game can be used as proposed in \cite{zhang2012physical}. The model is similar to that in \cite{zhang2010physical} in which the two sources, i.e., buyers, are the leaders, and the friendly jammers are followers, i.e., sellers. The sources determine their optimal FJ powers, and then the friendly jammers determine optimal power prices to maximize their own payments. The optimal solutions were obtained in a similar way as in Section~\ref{sec:Survey_eavesdropper_attack_external_Stackelberg}. It was proved that in the special case with one friendly jammer located close to the untrusted relay, the optimal power is monotonically decreasing and convex in the price when the other friendly jammers' prices are fixed. Therefore, there exists a unique Stackelberg equilibrium which is the pair of the optimal FJ power and price. However, a more general model with multiple untrusted relays can be considered to be an extension.  

\subsubsection{ACA-A}
\label{sec:Survey_eavesdropper_attack_internal_ACA_A}
To maximize secrecy rates for the sources while guaranteeing the cheat-proof property of the FJ power allocation, the authors in \cite{wang2013physical} employed the ACA-A approach \cite{zhang2013ascending}. The model is shown in Fig.~\ref{eavesdropper_ACA_A_multi_source_relay} with multiple source-destination pairs, one untrusted relay, i.e., the eavesdropper, and one friendly jammer. The source-destination pairs act as bidders which bid for the FJ power from the friendly jammer as the auctioneer, i.e., the seller. All steps of the power allocation process are implemented similar to those of the ACA-A approach, but the utility of the bidder is source-destination pair's utility as defined in \cite{zhang2010physical}. 

The ACA-A approach can also be found in \cite{wang2013joint}, but the FJ power is employed from the untrusted relay itself, i.e., the eavesdropper, to maximize the secrecy rate in a cooperative OFDMA network. The model includes multiple source-destination pairs as bidders and one untrusted relay as the auctioneer, i.e., the seller. Each bidder is assigned with a subcarrier, and the problem is to allocate the FJ power to each subcarrier to maximize the secrecy rate of the information communication on the subcarrier. Each bidder first defines its utility which is the difference between its secrecy rate and the price that it pays the relay. The bidders then submit FJ power requests to the relay to maximize the bidders' utilities. The optimal powers for the bidders are then obtained by using the ACA-A algorithm in \cite{wang2013physical}. The simulation results showed that the proposed scheme can improve the system sum secrecy rate around 5\% compared with the Stackelberg game-based approach \cite{han2009physical}. However, since the relay is untrusted itself, it can falsely report some information, e.g., the maximum power, which degrades the sum secrecy rate. 

\textbf{Summary:} This section discusses the applications of economic and pricing models to enhance the secrecy capacity of wireless communications. The reviewed approaches focus on how to impair the eavesdropper's ability to decode the information from sources to destinations. The reviewed approaches along with their references are summarized in Table~\ref{table_eavesdropper}. From the table, we observe that the auctions and Stackelberg game are mainly used for improving the secrecy capacity. Besides, the approaches against the external eavesdropper gain more attentions than the other types. However, eavesdroppers mentioned in this section are all passive, meaning that they only monitor and do not interfere communication channels. The next section reviews the applications of economic and pricing models to address the Denial-of-Service (DoS) attack which interrupts or suspends network services.

\begin{table*}
\caption{Applications of economic and pricing models for securing wireless communication against eavesdropping attack (CN: Cellular Network, CRN: Cognitive Radio Network, CWN: Cooperative Wireless Network, MANET: Mobile Ad hoc NETwork, WRN: Wireless Relay Network)}
\label{table_eavesdropper}
\scriptsize
\begin{centering}
\begin{tabular}{|>{\centering\arraybackslash}m{0.2cm}|>{\centering\arraybackslash}m{0.4cm}|>{\centering\arraybackslash}m{1.4cm}|>{\centering\arraybackslash}m{0.9cm}|>{\centering\arraybackslash}m{0.9cm}|>{\centering\arraybackslash}m{1.1cm}|>{\centering\arraybackslash}m{5.3cm}|>{\centering\arraybackslash}m{2.4cm}|>{\centering\arraybackslash}m{1.2cm}|>{\centering\arraybackslash}m{0.8cm}|}
\hline
\multirow{2}{*} {\textbf{}} & \multirow{2}{*} {\textbf{Ref.}} & \multirow{2}{*} {\textbf{Pricing model}} & \multicolumn{3}{c|} {\textbf{Market structure}} & \multirow{2}{*} {\textbf{Mechanism}} & \multirow{2}{*} {\textbf{Objective}} & \multirow{2}{*} {\textbf{Solution}} & \multirow{2}{*} {\textbf{Network}} \tabularnewline
\cline{4-6}
 & & & \textbf{Seller} & \textbf{Buyer} & \textbf{Item} & & &&\tabularnewline
\hline
\hline
\parbox[t]{2mm}{\multirow{9}{*}{\rotatebox[origin=c]{90}{ \hspace{-15cm} External eavesdropper}}}
& \cite{zhu2009improved}& Share auction& Friendly jammer&Sources& FJ power&Seller allocates the FJ power proportionally to the buyers' demands. &Secrecy capacity maximization&Nash equilibrium& MANET \tabularnewline \cline{2-10}

& \cite{zhang2013ascending}& ACA-T and ACA-A & Friendly jammer&Sources& FJ power&\textbf{ACA-T:} same as \cite{zhu2009improved}, but the seller adjusts the power price until the total demand is smaller than its maximum power. 

\textbf{ACA-A:} same as the ACA-T but the seller calculates the cumulative clinch. & Secrecy capacity maximization, allocation efficiency, and cheat-proofness&Walrasian equilibrium& CWN \tabularnewline \cline{2-10}

& \cite{li2015secure}& ACA-A & Base station&Mobile users& Subcarriers and FJ power& Same as the ACA-A in \cite{zhang2013ascending}, but the seller uses the Lyapunov's theorem to determine the optimal solutions of subcarrier, FJ power, and prices. & Secrecy capacity maximization, allocation efficiency, and cheat-proofness&Walrasian equilibrium & CN\tabularnewline \cline{2-10}

& \cite{wang2013stackelberg}  \cite{yue2012fairness}&Stackelberg game& Friendly jammer&Sources& FJ power& Buyers determine their FJ power demands, and then the seller determines the power price. & Utility improvement for both buyers and seller&Stackelberg equilibrium& CN \tabularnewline \cline{2-10}

& \cite{han2009physical}&Stackelberg game& Friendly jammers&Source& FJ power&Same as \cite{wang2013stackelberg}, but the buyer determines its FJ power for each seller.& Utility improvement for both buyer and sellers&Stackelberg equilibrium&MANET \tabularnewline \cline{2-10}

&\cite{maorui2011security}&Stackelberg game& Friendly relay&Source& Relay and FJ power&Buyer determines the relay and FJ power, and then the seller sets the optimal power price.& Utility improvement for both buyer and sellers&Stackelberg equilibrium & WRN\tabularnewline \cline{2-10}

&\cite{wang2012power}&Vickrey auction& Friendly relay&SU pairs& Relay and FJ power&Buyers submit their marginal secrecy rates as bids. Buyers with the highest bids are winners and pay the seller according to the Vickrey auction. & Secrecy rate improvement&Nash equilibrium& CRN \tabularnewline \cline{2-10}

&\cite{wang2011improve}&Sequential Vickrey auction& Friendly relay&SU pairs& Relay and FJ power&There are multiple rounds and in each round, the seller allocates a power unit to a buyer with the largest secrecy rate increase.  & Secrecy rate improvement, allocation efficiency&Subgame perfect equilibrium& CRN\tabularnewline \cline{2-10}

&\cite{deng2013truthful} \cite{liu2011enforce}&VCG auction& Friendly relays&Sources& FJ power&Buyer selects a number of sellers which maximize the sum of secrecy rates for the buyer. The payment is based on the VCG auction. & Secrecy rate maximization, truthfulness, individual rationality, and balanced budget&Bayesian Nash equilibrium& CWN \tabularnewline \cline{2-10}

&\cite{stanojev2011cooperative} &Vickrey auction& Source&Friendly jammers& Bandwidth and FJ power&Seller selects a buyer which maximizes its secrecy rate. The winner obtains the bandwidth and pays the seller its FJ power. & Average secrecy rate improvement, trustfulness&Nash equilibrium &MANET\tabularnewline \cline{2-10}

&\cite{ma2016improving} &Vickrey auction& PU&SUs& Time slot and FJ power&Same as \cite{stanojev2011cooperative}, but the winner gets the time slot from the seller.& Secrecy rate improvement, trustfulness&Nash equilibrium & CRN\tabularnewline \cline{2-10}


&\cite{bayat2012distributed}&Matching theory&Friendly jammers  &Sources&FJ power&Based on prices offered by the sellers, each buyer bids for its seller which maximizes its utility. The seller can adjust price to gain its utility. &Average secrecy rate improvement, social welfare maximization &Competitive equilibrium&MANET\tabularnewline \cline{2-10}

&\cite{wang2015dasi} \cite{wang2016auction}&Double auction&PUs &SUs&Bandwidth and FJ power& Each seller is matched with each buyer by using the maximum weighted matching algorithm and the double auction rule. & Truthfulness, individual rationality, and balanced budget &Competitive equilibrium& CRN\tabularnewline \cline{2-10}

&\cite{siyariprice} &Non-cooperative game&Power supplier &Sources&FJ power&Buyers obtain the optimal FJ power by using the KKT conditions. & Secrecy sum-rate improvement&Nash equilibrium& CN\tabularnewline \cline{2-10}

\hline
\parbox[t]{2mm}{\multirow{9}{*}{\rotatebox[origin=c]{90}{ \hspace{0cm} Internal eavesdropper}}}
&\cite{zhang2010physical} &Utility maximization&Friendly jammers &Sources&FJ power&Buyers obtain the optimal FJ power by solving the problem which is the maximization of their utilities.& Utility maximization for buyers&Optimal solution& WRN\tabularnewline \cline{2-10}

&\cite{zhang2012physical}&Stackelberg game&Friendly jammers &Sources&FJ power&Buyers determine the optimal FJ power, and then the sellers set the optimal prices.& Utility maximization for buyers and sellers&Stackelberg equilibrium&WRN\tabularnewline \cline{2-10}

&\cite{wang2013physical}&ACA-A&Friendly jammer&Source pairs&FJ power&Same as \cite{zhang2013ascending}.& Secrecy rate maximization&Walrasian equilibrium&WRN\tabularnewline \cline{2-10}

&\cite{wang2013joint}&ACA-A&Untrusted relay&Source-destination pairs&FJ power&Same as \cite{wang2013physical}, but the secrecy rate of a buyer is defined for each subcarrier which depends on the mutual information of private channels. & Sum-secrecy rate maximization&Walrasian equilibrium&WRN\tabularnewline \cline{2-10}

\hline
\end{tabular}
\par\end{centering}
\end{table*}

\begin{table*}[h]
\caption{A summary of advantages and disadvantages of major approaches for securing wireless communication against eavesdropping attack}
\label{table_sum_advantage_eavesdropping}
\scriptsize

\begin{centering}
\begin{tabular}{|>{\centering\arraybackslash}m{2cm}|>{\centering\arraybackslash}m{7.8cm}|>{\centering\arraybackslash}m{6cm}|}
\hline
\cellcolor{myblue} &\cellcolor{myblue} &\cellcolor{myblue} \tabularnewline
\cellcolor{myblue} \multirow{-2}{*} {\textbf{Major approaches}} &\cellcolor{myblue} \multirow{-2}{*} {\textbf{Advantages}} &\cellcolor{myblue} \multirow{-2}{*}{\textbf{Disadvantages}} \tabularnewline
\hline
\hline
\cite{zhu2009improved} &\begin{itemize} \item Do not require exact knowledge of channel information  \end{itemize} & \begin{itemize}  \item  Support only one eavesdropper \end{itemize}\tabularnewline \cline{2-3}
  \hline
\cite{zhang2013ascending} &  \begin{itemize} \item Be easy to be implemented and have fast convergence\end{itemize}& \begin{itemize} \item Support only one eavesdropper  \end{itemize}\tabularnewline \cline{2-3}
  \hline
\cite{wang2013stackelberg} & \begin{itemize} \item Achieve win-win solution and fast convergence \end{itemize} & \begin{itemize} \item Do not guarantee the fairness among sources \end{itemize} \tabularnewline \cline{2-3}
  \hline
\cite{wang2011improve}  &\begin{itemize} \item Achieve high power utilization and do not require external FJ jammers \end{itemize}&\begin{itemize} \item Support only one relay and have high complexity \end{itemize} \tabularnewline \cline{2-3}  
  \hline
\cite{deng2013truthful}  &\begin{itemize} \item Support multiple relays \end{itemize}&\begin{itemize} \item Do not achieve the balanced budget  \end{itemize} \tabularnewline \cline{2-3}  
  \hline
\cite{bayat2012distributed} &\begin{itemize} \item Support multiple friendly jammers and sources \end{itemize} &\begin{itemize} \item Have high complexity \end{itemize} \tabularnewline \cline{2-3}
\hline
\end{tabular}
\par\end{centering}
\end{table*}

\section{Denial-of-Service (DoS) attack}
\label{sec:Survey_DoS_attack}
A DoS attack is a deliberate action of one or many nodes with the aim of interrupting or degrading services of wireless networks. This section reviews applications of economic and pricing models for defending the different DoS attacks. The common DoS attacks include:
\begin{itemize}
\item \textit{Jamming attack:} This attack is executed by generating high-power signals to interfere the legitimate communication channels and thus reducing the SINR at the legitimate receivers. While attempting to cause the interference, the jammer is subject to the power price which is set by an energy supplier. Thus power pricing strategies can be applied to discourage jammers to spend power for their jamming actions.
\item \textit{DDoS attack:} The DDoS attack, also known as the botnet attack, uses the network resources of a large number of compromised nodes, i.e., bots, inside the network to encumber the legitimate accesses to the resources at a receiver \cite{zargar2013survey}. Pricing models as distributed solutions are efficiently used to prevent attack behaviors of the bots.
\item \textit{Black hole attack:} In this attack, a malicious node in a route, e.g., a relay node, silently drops or discards incoming packets from a source without forwarding them to the source's intended destination. Pricing strategies can be used to detect and isolate such malicious nodes.
\end{itemize}

\subsection{Jamming Attack}
\label{sec:Survey_DoS_attack_jamming}
 Anti-jamming techniques such as the frequency hopping \cite{popovski2006strategies} have been commonly used. However, such traditional techniques require a large number of spectrum resources. Due to the resource constraint, a more potential approach is to exploit the cooperation of legitimate users in the network. More specifically, legitimate users with good links can trade their capacities with users under the jammed links. 

\begin{figure}[t!]
 \centering
\includegraphics[width=6.7cm, height=5.5cm]{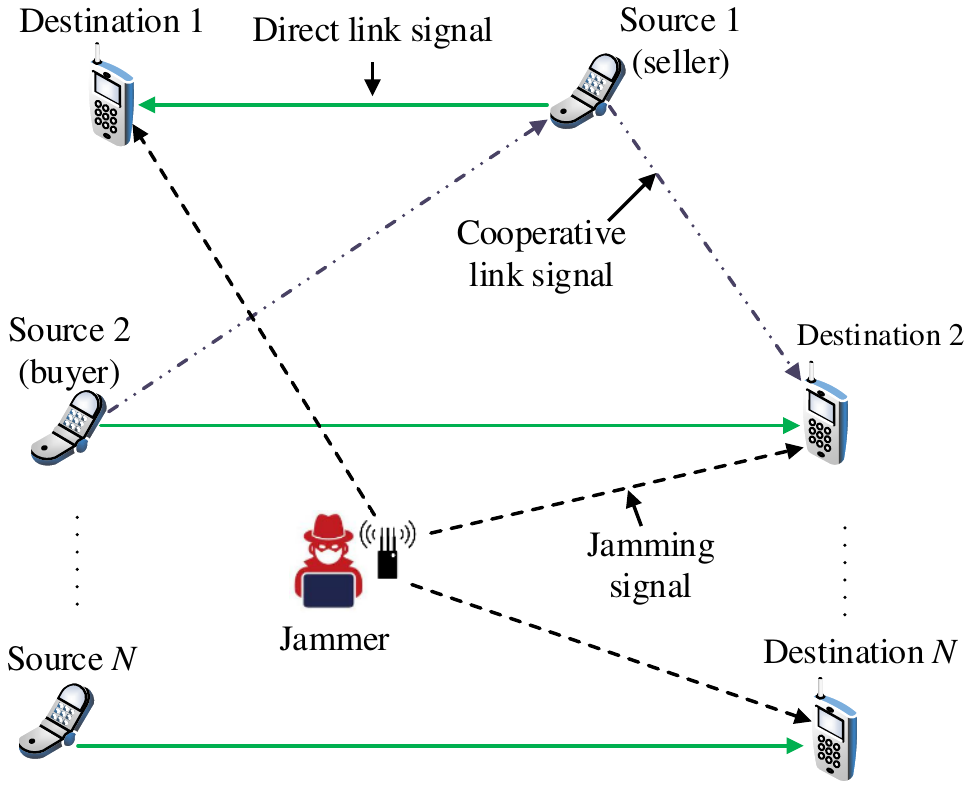}
 \caption{Anti-jamming technique based on the cooperation of legitimate users.}
 \label{anti_jamming_cooperative}
\end{figure}

Inspired by the aforementioned idea, the authors in \cite{zhang2014cooperative} proposed a cooperative anti-jamming scheme to optimize the fairness-constrained network throughput. The model is shown in Fig.~\ref{anti_jamming_cooperative} which includes one jammer and multiple legitimate users each consisting of a source-destination pair. The jammer degrades the throughput of the users by causing interference on their channels with corresponding jamming powers, i.e., the jamming profile. Each user has its own sensing probability profile which includes probability that the user senses the channels. The optimization problem is formulated to determine the users' sensing probability profiles to maximize the sum utility, i.e., expected capacity, of all the users given the jammer's jamming profile. The optimization problem is non-convex, and the iterative best-response algorithm based on a general pricing mechanism \cite{scutari2014decomposition} was used. At each iteration, the user maximizes its own utility minus a pricing term. The pricing term generally depends on the marginal decrease of the sum-utility of the other users due to the variation of the user's sensing probability profile. By using the second-order derivative, the user's utility was proved to be strongly concave with respect to its channel sensing profile. Then, based on the Descent Lemma in \cite{bertsekas1999nonlinear}, the algorithm was proved to converge to a feasible and stationary solution. The simulation results showed that the total utility of the proposed algorithm in \cite{zhang2014cooperative} gains up to 19.8\% improvement compared to that of the frequency hopping algorithm. 


In military environments, a jammer as an adversary can perform jamming and eavesdropping simultaneously. The authors in \cite{tang2016combating} enhanced the secrecy rate for one legitimate source-destination pair in presence of such an adversary. The secrecy rate is the capacity rate of the legitimate transmission channel minus that of the wiretap channel. The adversary causes interference to the destination while the source attempts to increase its transmit power to maintain the desired secrecy rate. The interaction between the adversary and the source is thus modeled by a game in which the adversary's strategy is to find the jamming power to maximize its utility. In fact, by consuming more power, the adversary is subject to the linear price of the power which is set by an energy supplier. Thus its utility is the difference between its capacity rate and the price that it pays. The utility is concave in the jamming power, and thus the optimal jamming power is determined by using the first-order derivative. The source determines its minimum transmit power subject to the desired secrecy rate based on the optimal jamming power. The pair of the optimal jamming power and transmit power thus forms a unique Nash equilibrium. The simulation results showed that the legitimate transmit power is always positively related to the jamming power. Also, the jamming power decreases rapidly with the increase of the power price. This naturally makes the adversary more conservative with its jamming action. 

\subsection{Distributed DoS (DDoS) attack}
\label{sec:Survey_DoS_attack_Botnet}
Compared with the jamming attack, the DDoS attack is more severe \cite{jensen2008impact} since it is launched by a large number of malicious nodes, i.e., compromised nodes or bots, which are distributed across the network. In models discussed in this section, the malicious nodes harm regular nodes by transmitting noises with the high power at a base station. The malicious nodes are also assumed to be rational. Since the malicious nodes need to use more network resources for the attack actions, they are subject to the resource price which is set by the base station. Thus pricing models can be used to set the amount of network resources according to the users' behaviors such that their malicious behaviors are prevented.

The authors in \cite{shen2016resource} addressed the DDoS attack in a multiple-access system with the coexistence of malicious users, i.e., compromised users, and selfish users. The model is shown in Fig.~\ref{DDoS_attack_for_intro} in which a base station as a seller allocates the power to users (the malicious and selfish users) as buyers to satisfy their rate-based QoS requirements. Since the base station does not use the successive interference cancellation, each user suffers from the interference of all the other users. Thus the interaction among them can be modeled by the non-cooperative game in which each user's strategy is to determine the allocated power to maximize its utility. The utility of the user is a function of its capacity rate, the demand of other users, the individual price that it pays the base station, and its \textit{private type}. In particular, the user's private type indicates the extent of its behavior \cite{chorppath2015adversarial}, and the individual price is inversely proportional to the user's private type. Such a pricing strategy discourages malicious users to spend much power to harm the other users. Using the Crammer's rule \cite{gong2002note}, the best response power allocation for users is determined and converges to a unique Nash equilibrium. However, how to learn the users' private types was not specified. 


To learn private types of users, the authors in \cite{chorppath2014bayesian} assumed that the base station can observe the users in a sufficient time to get probabilistic information of their behaviors. The base station's problem is to find the resource allocation that maximizes the users' social welfare, i.e., the sum of utilities, while preventing the maliciousness of users. Here, the utilities of malicious users are multiplied with factors related to their probabilistic information, namely \textit{degree of maliciousness}. The utility of each user is strictly concave in its allocated power. The Lagrange multiplier method is then used to solve the optimization problem with their interpretations as powers and prices that the users are willing to pay. At each iteration, the power prices are updated by using the gradient projection method, and the users determine their optimal power. It was proved that there exists a unique optimal solution for the allocated power and the prices. It is also worth noting that the optimal price for each user is proportional to its probability belief, i.e., the probability that the user performs a malicious action. This means that the price for the users is higher with higher probability belief. Thus they are incentivized to not harm other users to avoid high cost.  

It can be seen that the pricing strategy in \cite{chorppath2014bayesian} is a type of the Bayesian pricing \cite{segal2003optimal} in which prices for users are set based on the probabilistic assumptions on their behaviors. This pricing strategy is also similar to the differential pricing scheme \cite{varian2014intermediate} which sets different prices for different users according to their behaviors. However, the differential pricing scheme aims to maximize the profit of the base station rather than maximizing the users' social welfare. Generally, the challenge of such pricing mechanisms is the inaccuracy of estimating the users' behaviors, especially when the observation time is insufficient. The Gaussian process regression learning technique using training data sets \cite{chorppath2011learning} can be applied to improve the estimation accuracy as proposed in \cite{chorppath2015bayesian}. Moreover, the learning technique also allows the base station to infer the utilities of both regular and malicious users. As shown in the simulation results, the estimated utilities of users are almost the same as their actual utilities. In other words, the users' utilities are assumed to be known at the base station. Thus the centralized auction mechanisms can be adopted in which the base station makes centralized decisions on both the power level and price for all users as proposed in \cite{chorppathbayesian}. 

The model in \cite{chorppathbayesian} is similar to that in \cite{chorppath2014bayesian}. The users as bidders submit bids including their optimal power requests to the base station as the seller. The optimal power requests of users are determined similar to those in \cite{chorppath2014bayesian}. Upon receiving the bids, the base station allocates power to users according to the share auction, i.e., the proportional allocation, as presented in \cite{zhu2009improved}. However, the price that the user pays is proportional to its power request and inversely proportional to the probability of the user being malicious. This payment scheme will force the malicious users to act as regular users in the network. The simulation results showed that the additional cost for the malicious user decreases when the probability that the user performs a malicious action decreases. However, in practice, the malicious users may continuously observe the network and update their degree of maliciousness to maximize the harm to the regular users while minimizing the detection probability. 

\subsection{Black Hole Attack}
\label{sec:Survey_DoS_attack_Hole}
A black hole problem is actually a misbehavior routing problem \cite{tseng2011survey} in which one malicious node uses the same routing protocol as the regular nodes, but it drops the routed packets and does not forward the packets to its neighbors. The traditional approaches using the punishment mechanism such as CONFIDANT protocol \cite{buchegger2002performance} were proposed to assure routing security and fairness. However, the approaches still cannot eliminate the malicious nodes completely. By observing that a selfish node in the network always desires to receive the higher reputation than other nodes, an efficient solution is to make the node compete for the reputation with others through forwarding incoming packets. As a result, competitive approaches such as auction or non-cooperative game are used.

\begin{figure}[t!]
 \centering
\includegraphics[width=6.9cm, height=4.3cm]{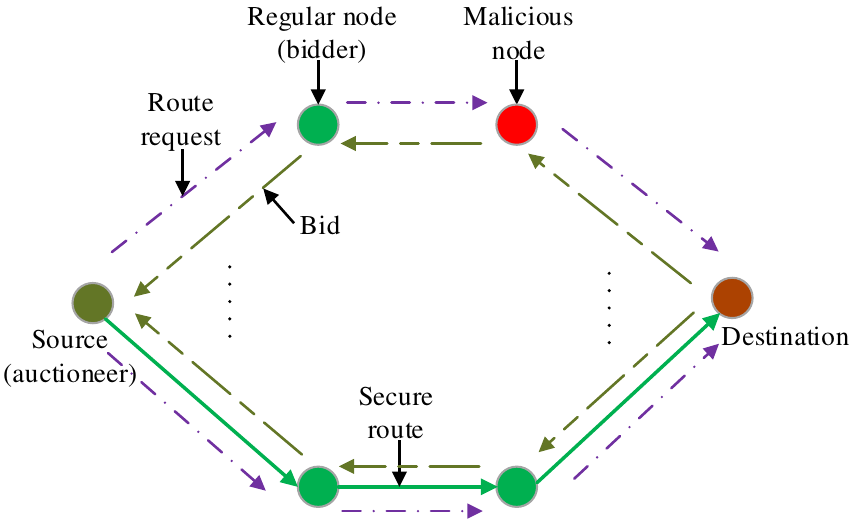}
 \caption{Secure routing protocol based on auction.}
 \label{DoS_first_price_auction}
\end{figure}

The authors in \cite{agah2006security} investigated a secure sensor network
routing protocol by using the first-price sealed-bid auction to
eliminate malicious sensors. This auction is used since it requires little communication or interaction among the sensors. A source sensor as a seller selects a secure route which is formed by other sensors, i.e., bidders, to forward packets to its destination. The secure route is required to not include any malicious node that attempts to drop incoming packets. First, the source sensor broadcasts a route request to its
neighbors as shown in Fig.~\ref{DoS_first_price_auction}. Then, the neighbors forward
this request to others. The destination may receive several requests and confirms with the source by sending a reply message including the bids. Each bid represents a route and contains the sum of utilities of all sensors on the route. Here, the utility of a sensor is a function of its remaining power and reputation. The source selects the route which has the highest
bid as the winner for its data transmission. The sensors
on the winning route pay the source sensor a percentage of the initial power and receive a good reputation as the reward. Since the sensors on a route want to receive the high reputation from the source sensor, they are motivated to cooperate to forward incoming packets.

The approach in \cite{agah2006security} can mitigate the malicious sensors causing the disruption to routing. However, it does not consider the scenario in which some sensors could agree to join the auction but will subvert the route later. A \textit{watch-list} \cite{agah2005enforcing} which records the sensors' misbehavior can
be used to recognize such malicious sensors and then notify all other
sensors to not communicate with them. The simulation results in \cite{agah2006security} showed that the total
number of dropped packets of the proposed scheme is two-thirds less than that of the CONFIDANT protocol \cite{buchegger2002performance}. The reason is that sensors with bad reputation in the proposed scheme are ignored by the majority of sensors. However, both proposed schemes, i.e., \cite{agah2006security} and \cite{agah2005enforcing}, do not consider the case in which some sensors agree not to outbid each other. This scenario has the overall effect of lowering the winning bid and increasing the number of dropped packets. 


Apart from the first-price sealed-bid auction, the Dutch auction (as shown in Section~\ref{subsec:Game_theory_auction}) can also be used to address the back hole attack as proposed in \cite{reidt2009fable}. The Dutch auction is an appropriate solution since it can (i) provide users sufficient incentives due to its payment policy and (ii) guarantee that the malicious nodes can be quickly revoked due to its simplicity. The model consists of honest nodes as bidders, a trust authority as the auctioneer, and a malicious node as the item. The bidders use Intrusion Detection Systems (IDSs) to collect evidence of malicious behavior, i.e., the non-forwarding of packets, of suspicious neighbor nodes. Each bidder has its own private value for the suspicious node which is the \textit{risk appetite} to revoke the suspicious node. The risk appetite is defined as \textit{``1-desired certainty in revoking the suspicious node''} and depends on the accuracy of the IDS. The bidder which takes the highest risk and revokes the suspicious node earliest is known as the winner and receives a reward for its effort by the trust authority. The winner's profit is the difference between its private value and the price quoted in the auction. The simulation results showed that when the IDS has a higher detection probability, the malicious node is revoked after fewer auction rounds. However, the probability of revoking the malicious node significantly lowers when the number of bidders decreases since the competition among bidders is low. 

The Dutch auction can be combined with the reverse auction to enforce the cooperation among nodes in a MANET as proposed in \cite{chatterjee2012stacrp}. In the network model under consideration, a pair of source and destination nodes are buyers, and intermediate nodes in the network are sellers. The source sends a routing request to the intermediate nodes to forward its packet to the destination node. Upon receiving the routing request from the source node, the intermediate nodes reply with their prices. The source node uses the reverse auction to select an intermediate node with the lowest price as the forwarding node. If the forwarding node is not in the list of the neighbors of the destination node, the forwarding node adopts the Dutch auction to sell the packet forwarding service to the next hop node. Note that the proposed scheme uses the Markov chain to predict the trust of nodes in the network. Thus the source and forwarding nodes can check the trust levels of their neighbor nodes. Finally, the total cost is paid by both the source and destination nodes. This enforces both of them to behave truthfully and prevent them from generating false routing request flooding. To provide incentives, the forwarding node receives the virtual currency which can be used to forward its own packet in future. Simultaneously, the trust level of the forwarding node is increased. 

The proposed schemes in \cite{reidt2009fable} and \cite{chatterjee2012stacrp} require centralized entities, e.g., the trust authority, which may not be available in infrastructure-less networks such as MANETs. Fully distributed schemes using the mean field game theory can be applied as proposed in \cite{wang2014mean}. Such a scheme enables each node to make defense strategies based on only its own state information, i.e., a combination of its energy and information assets, and the aggregate effect of the other nodes. 

\textbf{Summary:} This section discusses the applications of economic and pricing models for the DoS attack prevention. The reviewed approaches aim to provide resource pricing mechanisms such that the users do not have any incentive to perform their attack actions. We have considered three common types of DoS attacks. The reviewed approaches along with their references are summarized in Table~\ref{table_DoS_attack}. As seen, the DDoS attacks have received more attentions because they are more serious and common than the other types of attacks. However, the pricing models developed for preventing the DoS attacks in overall are relatively few. The next section reviews the applications of economic and pricing models to enhance the information security, i.e., privacy preserving and integrity.

\begin{table*}
\caption{Applications of economic and pricing models for defending against DoS attacks (CN: Cellular Network, MANET: Mobile Ad hoc NETwork, WSN: Wireless Sensor Network)}
\label{table_DoS_attack}
\scriptsize
\begin{centering}
\begin{tabular}{|>{\centering\arraybackslash}m{0.2cm}|>{\centering\arraybackslash}m{0.4cm}|>{\centering\arraybackslash}m{1.3cm}|>{\centering\arraybackslash}m{0.9cm}|>{\centering\arraybackslash}m{1cm}|>{\centering\arraybackslash}m{0.9cm}|>{\centering\arraybackslash}m{5cm}|>{\centering\arraybackslash}m{2.8cm}|>{\centering\arraybackslash}m{1.2cm}|>{\centering\arraybackslash}m{0.8cm}|}
\hline
\multirow{2}{*} {\textbf{}} & \multirow{2}{*} {\textbf{Ref.}} & \multirow{2}{*} {\textbf{Pricing model}} & \multicolumn{3}{c|} {\textbf{Market structure}} & \multirow{2}{*} {\textbf{Mechanism}} & \multirow{2}{*} {\textbf{Objective}} & \multirow{2}{*} {\textbf{Solution}}& \multirow{2}{*} {\textbf{Network}} \tabularnewline
\cline{4-6}
 & & & \textbf{Seller} & \textbf{Buyer} & \textbf{Item} & & &&\tabularnewline
\hline
\hline
\parbox[t]{2mm}{\multirow{9}{*}{\rotatebox[origin=c]{90}{\hspace{1.5cm} Jamming attack}}}
&  \cite{zhang2014cooperative}&General pricing &Legitimate users &Legitimate users& Free-jamming links& Each buyer pays the seller with a price based on the marginal decrease of the sum-utility of the other buyers due to the buyer's strategy.& Total utility improvement, and network traffic increase&A feasible and stationary solution &MANET\tabularnewline \cline{2-10} 

&\cite{tang2016combating}&Linear pricing &Energy supplier&Adversary& Power& Given the power price, the buyer determines its optimal power, and then the source finds its transmit power for the secrecy rate guarantee.& Secrecy rate guarantee, and energy efficiency&Nash equilibrium &MANET\tabularnewline \cline{2-10} 
\hline
\parbox[t]{2mm}{\multirow{9}{*}{\rotatebox[origin=c]{90}{\hspace{-1cm} Distributed DoS attack}}}
& \cite{shen2016resource}&Non-cooperative game &Base station&Users& Power& The best response power allocation for buyers depends on their private types and is obtained by using the Crammer's rule.&  PoM reduction, and rate-QoS satisfaction&Nash equilibrium & CN\tabularnewline \cline{2-10} 

& \cite{chorppath2014bayesian}&Utility maximization &Base station&Users& Power& Based on the probabilistic information of buyers' behaviors, the seller finds the optimal power allocation for the buyers using the Lagrange
multiplier method.& PoM reduction, and buyes' utility maximization&Optimal solution & CN\tabularnewline \cline{2-10} 

& \cite{rasmussen2004gaussian}&Utility maximization &Base station&Users& Power& Same as \cite{chorppath2014bayesian}, but the Gaussian process regression learning technique is used.& PoM reduction, and buyes' utility maximization&Optimal solution & CN\tabularnewline \cline{2-10} 

& \cite{chorppathbayesian}&Share auction &Base station&Users& Power&Seller allocates the power to buyers according to the share auction in \cite{zhu2009improved}. Each buyer pays the seller the price which is inversely proportional to the probability of the buyer being malicious. & Malicious behavior minimization, and buyes' utility maximization&Bayesian Nash equilibrium & CN\tabularnewline \cline{2-10} 

\hline
\parbox[t]{2mm}{\multirow{9}{*}{\rotatebox[origin=c]{90}{\hspace{0cm} Black hole attack}}}

&  \cite{agah2006security}& First-price sealed-bid auction& Source sensor&Forwarding sensors& Reputation&Seller selects the route with the highest total utility of buyers as the winner.  & Secure routing improvement&Nash equilibrium & WSN\tabularnewline \cline{2-10} 

& \cite{agah2005enforcing}& First-price sealed-bid auction& Source sensor&Forwarding sensors& Reputation&Same as \cite{agah2006security}, but the watch-list is used to monitor and update the misbehavior of buyers. & Secure routing improvement&Nash equilibrium &WSN\tabularnewline \cline{2-10} 

&\cite{reidt2009fable}&  Dutch auction&Trust authority&Honest nodes& Malicious node&The buyer which takes the highest risk appetite and revokes the suspicious node earliest is the winner and gets a reward. & Optimal revocation of malicious node&Nash equilibrium &MANET\tabularnewline \cline{2-10} 
\hline
\end{tabular}
\par\end{centering}
\end{table*}

\begin{table*}[h]
\caption{A summary of advantages and disadvantages of major approaches for defending against DoS attacks}
\label{table_sum_advantage_DoS}
\scriptsize
\begin{centering}
\begin{tabular}{|>{\centering\arraybackslash}m{2cm}|>{\centering\arraybackslash}m{8.5cm}|>{\centering\arraybackslash}m{6cm}|}
\hline
\cellcolor{myblue} &\cellcolor{myblue} &\cellcolor{myblue} \tabularnewline
\cellcolor{myblue} \multirow{-2}{*} {\textbf{Major approaches}} &\cellcolor{myblue} \multirow{-2}{*} {\textbf{Advantages}} &\cellcolor{myblue} \multirow{-2}{*}{\textbf{Disadvantages}} \tabularnewline
\hline
\hline
 \cite{zhang2014cooperative}&\begin{itemize} \item Provide a distributed solution  \end{itemize} & \begin{itemize}  \item  Support only one jammer\end{itemize}\tabularnewline \cline{2-3}
  \hline
\cite{shen2016resource} &  \begin{itemize} \item Provide a distributed solution and support multiple malicious and selfish users \end{itemize}& \begin{itemize} \item Require learning techniques  \end{itemize}\tabularnewline \cline{2-3}
  \hline
 \cite{chorppathbayesian} & \begin{itemize} \item Achieve centralized decisions on both power and price \end{itemize} & \begin{itemize} \item Be challenging to infer the utilities of users \end{itemize} \tabularnewline \cline{2-3}
  \hline
\cite{agah2006security} &\begin{itemize} \item Have low overhead communication \end{itemize}&\begin{itemize} \item Do not record misbehaviors of malicious sensors  \end{itemize} \tabularnewline \cline{2-3}  
  \hline
\cite{reidt2009fable}& \begin{itemize} \item Be simple to be implemented and revoke quickly malicious nodes \end{itemize}& \begin{itemize} \item Require IDSs and the trust authority  \end{itemize} \tabularnewline \cline{2-3}
\hline
\end{tabular}
\par\end{centering}
\end{table*}

\section{Privacy and confidentiality}
\label{sec:Survey_privacy_concerns}
 As discussed in Section~\ref{sec:Survey_eavesdropper_attack}, attackers are easier to perform monitoring and eavesdropping communication links between regular users if the attackers have full knowledge of the locations of the regular users in the network. In realistic environments, when the locations of legitimate users are revealed, they would have a high risk from physical attacks by the adversaries. Thus it is of the great importance to investigate the location privacy which refers to the ability of preventing the adversaries from learning the regular users' current or past locations \cite{liu2007protecting}. This section reviews economic and pricing approaches to address the user information privacy leakage issue. In particular for the wireless environments, the two following scenarios are considered. 
 
\begin{itemize}
\item \textit{Privacy leakage in sensing data collection:} In the sensing data collection, sensors or phone users perform sensing in an area of interest and submit the data to the central controllers for further processing. However, the collected sensing data may include location information which reveals sensitive information, e.g., home address and identifiers of the users. Economic and pricing models are developed and integrated with cryptographic algorithms to obtain sensing data with high quality and low cost while protecting the users' privacy \cite{luong2016data}.
\item \textit{Privacy leakage in spectrum allocation:} In the spectrum allocation, users which are located far enough from each other can be allocated the same spectrum without causing interference with each other. Thus they are required to submit their location information when requesting the spectrum. This naturally reveals the users' location information. The combination of pricing models and cryptographic algorithms can support efficient and privacy-preserving spectrum allocation for the users. 
\end{itemize}
\subsection{Privacy leakage in sensing data collection}
\label{sec:Privacy_preserving_data_collection} 


A sensing data collection process can be implemented through a WSN or a Mobile Crowdsensing Network (MCN). Both of them perform gathering sensing data from interest areas. However, the WSN is typically composed of a large number of autonomous and resource-limited sensors to collect the data. On the contrary, the MCN explores the data from mobile devices, e.g., smartphones, of users \cite{luong2016data}. As a result, the security issue in the WSN is typically the misbehavior routing such as black hole attack while that in the MCN is the physical attack due to the users' privacy leakage. The black hole attack in the WSN was addressed by pricing models combined with misbehavior detection schemes as presented in Section~\ref{sec:Survey_DoS_attack_Hole}. This section discusses pricing models integrated with cryptographic method to prevent the privacy leakage for users in the MCN.

\begin{figure}[t!]
 \centering
\includegraphics[width=4.8cm, height=6cm]{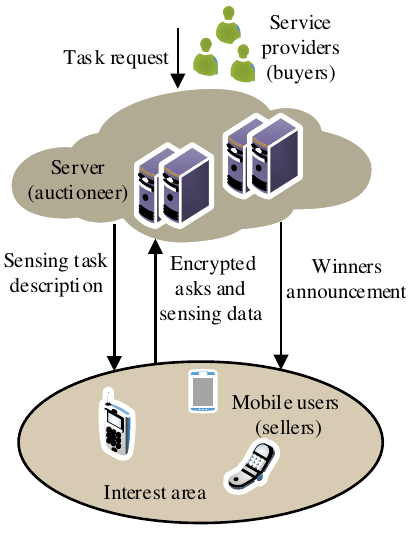}
 \caption{Sensing data collection based on the reverse Vickrey auction.}
 \label{data_aggregation_sealed_bid_reverse_auction}
\end{figure}

A typical data aggregation model in the MCN is shown in Fig.~\ref{data_aggregation_sealed_bid_reverse_auction} which consists of a server and participants, i.e., mobile users. One of the most common approaches used in the model is the reverse auction \cite{luong2016data}. The main motivations of using such auction are to (i) reduce the incentive cost, (ii) achieve high quality data, (iii) prevent bidders from knowing bids of each other through simultaneous sealed bid submission, and (iv) provide incentives to the users for contributing their data. The basic data aggregation based on the sealed-bid reverse auction works as follows. The server as an auctioneer first broadcasts the sensing task description from customers to all mobile users, i.e., sellers. Interested users accept and perform the sensing task. Upon completing the sensing task, the users submit their asking prices including the sensing data and the prices to the server. The server selects the users with the lowest asking prices as the winners and makes the payments to them. 

In fact, the auctioneer selects the winners based on not only their asking prices, but also the areas where the users perform sensing tasks. For example, if users are closer to the area of interest, their sensing data is likely to be more useful and thus has higher utility. As such, the users have more opportunities to become the winners. In this scenario, the users are often required to specify their locations along with the asking prices. This leads to the location information leakage of the users. Even worse, there may be attackers in the network which harm the users. To preserve the location information, users only submit their asking prices with obscure locations as proposed in \cite{singla2013incentives}. After the asking process, the winning users reveal their actual locations when submitting the sensing data. However, utility of the auctioneer in this case can be hampered since it cannot evaluate the actual utility of the sensing data without having the knowledge of users' accurate locations. This makes the auctioneer overpay or underpay for the received sensing data. 

In practice, even if the users' locations are obscure, the attacker or auctioneer can infer the users' locations from the asking prices. Indeed, a user closer to the area of interest usually evaluates its sensing data at high utility, thus setting the higher asking price than that of the others. From the asking price together with the knowledge of interest area, the attacker or auctioneer can infer the user's location. In \cite{liu2017invisible}, the auctioneer broadcasts the payment for sensing data that are collected at a specific location. Based on the payment information, each participant decides whether to upload the data or not. There is no asking price submission, and this reduces the risk of privacy leakage to the attacker. However, the asking price is still revealed to the auctioneer. Therefore, the price privacy needs to be preserved. In \cite{sun2014privacy}, the asks are encrypted before being submitted using the Time Lapse Cryptography (TLC) which provides the computation efficiency and correctness. Moreover, before being transmitted to the auctioneer, the asks are signed electronically by the users by employing the Nyberg-Rueppel signature scheme \cite{camenisch1995blind} to keep the confidentiality of the asks. The auctioneer determines the winners based on the encrypted asks rather than their real ones. Specifically, the users which have the encrypted asks lower than a payment threshold are selected as the winners to provide sensing data. The payment threshold influences the outcome of the auction. However, how to determine the threshold was not explained. 

The same approach can be found in \cite{wang2016incentive}, but a general auction was adopted to provide online privacy-aware incentives to users. In this model, users as sellers come and submit their asks for a sensing task sequentially requested by a platform, i.e., the buyer. First, using TLC, each user encrypts its asking price, sensing time, and pseudonym with the platform's public key. The user then makes and signs a commitment and sends its bidding request to the platform. Given the requests, the platform determines the winner based on the user's past reputation and marginal utility, i.e., the utility gained from using the sensing data, by applying the platform decryption key. A reject or accept decision is encrypted and signed before being sent to the users. If a user is accepted, it starts its sensing task. The sensing data is encrypted and signed before being submitted to the platform. If the data is trustable, the platform increases the reputation for the user as a reward. Otherwise, the platform decreases the user's reputation as the punishment. The reputation punishment makes the proposed mechanism truthful. As shown in the simulation results, the number of trustful users can get to a stable status after 10 transactions, each of which is defined as the interaction between the user and the platform in a task. However, some users can accumulate reputations in some transactions and then behave unreliably in later transactions. Thus the malicious fluctuation problem of users needs to be further investigated.

The general auction was also adopted in \cite{wu2016magicrowd} to provide privacy-aware incentives and fairness for the users. The model consists of a campaigner as a buyer, an auctioneer, and users as sellers. Each user submits an ask to the auctioneer including its asking price, true value, and location information. The auctioneer clusters the users into crowds, each of which satisfies the \textit{spatial-price approximation}, i.e., the total difference in asking prices and locations among users in the same crowd is minimum, and the $k$-\textit{anonymity} location privacy, i.e., a user cannot be distinguished from the location information of at least $(k-1)$ other users in the same crowd. The auctioneer then employs the bargaining game \cite{muthoo1999bargaining} to bargain with the campaigner on a deal price for each crowd. The objective aims at maximizing the product of the crowd's and the campaigner's utility gains. The auctioneer allocates the deal price to the users. The payment for each user is inversely proportional to the difference between its asking price and its true value. Such a payment encourages the users to submit asking prices close to their true valuation. As shown in the simulation results, although users in the same crowd have different asking prices, their payments are almost the same, i.e., the fairness among the users is achieved. 


Similar to \cite{wu2016magicrowd}, the user anonymity and fairness issues were also considered in \cite{shi2013sealed}, but the sealed-bid multi-attribute reverse auction was employed. The sealed-bid multi-attribute reverse auction is similar to the reserve auction, but the former allows the auctioneer to evaluate an ask based on more attributes in addition to the price. More specifically, the attributes of an ask are the price, the location accuracy, and the sampling frequency. These attributes are converted to the ask utility score through the linear utility function used in \cite{bichler2005configurable}. The asks are encrypted by using the Paillier cryptosystem \cite{paillier1999public} before being submitted to the auctioneer. The auctioneer then adopts the private set intersection algorithm \cite{freedman2004efficient} for determining the winner. The proposed algorithm does not need to open asks while still achieving the winner determination and the public verifiability. The public verifiability is provided by the use of the bulletin board which allows anybody to check the identities of sellers and confirm whether their asks are valid or not. Thus the strong ask privacy is maintained during the auction. According to the security analysis, the proposed scheme satisfies the security requirements of an electronic auction, e.g., anonymity, ask privacy, and public verifiability. 

The same approaches can be found in \cite{dimitriou2015privacy}. However, instead of using the Paillier cryptosystem, the authors in \cite{dimitriou2015privacy} masked each seller's ask with a hash value which is computed from a random number. The one-wayness property of the hash function ensures that the sellers' asks remain hidden. This eliminates the possibility that a seller or the auctioneer can divulge information about the others throughout the asking process. As a result, the confidentiality of asks is better maintained compared to \cite{shi2013sealed}. The winners are then determined similar to \cite{shi2013sealed}. Note that to increase the winning opportunities for the losers in the next rounds, the proposed scheme in \cite{dimitriou2015privacy} includes the number of previously lost auction rounds as one attribute when evaluating the asks. 

Different from \cite{dimitriou2015privacy}, the authors in \cite{li2017scalable} used a key generator which randomly generates and distributes a series of polynomial values and IDs for the participants. Then, by leveraging the tools such as the Lagrange polynomial interpolation and fixed point representation, the auctioneer determines the winners and their corresponding payments without leaking the ask privacy of participants to any of the other parts including itself. Note that the payments for the winners are based on the VCG payment policy to guarantee the truthfulness of the asks. 

In addition to the sellers' privacy preserving, the authors in \cite{holzbauer2012socially} considered the power consumption and the history participation of the sellers. The proposed approach is essentially the first-price sealed-bid reverse auction. However, the ask evaluation is similar to the sealed-bid multi-attribute reverse auction. Accordingly, once receiving the sensing data requirement from the auctioneer, i.e., a data sink, sellers generate their own asks. The seller's ask is calculated based on its true value, its current power level, the number of auction rounds that the seller has participated. In particular, if the seller performs sensing frequently, i.e., the large number of auction rounds, the seller should receive a higher incentive to compensate the loss of the seller's location information associated with its data transfered to the sink. Based on the asks, the sink selects a predefined number of sellers with the lowest asks as the winners. The simulation results showed that when the predefined number is very high, i.e., 100, the average inter-win bid time of the proposed scheme, which is defined as the average number of ticks that elapsed between two wins for a given node, of a seller is low since the sellers with low privacy concern are selected frequently. Whereas this value of the baseline \cite{lee2010sell} changes only slightly which implies that the baseline cannot be calibrated to meet different privacy requirements.

Apart from the privacy concerns in crowdsensing networks, the privacy for drivers in Vehicular Cloud Computing (VCC) has recently been considered. The VCC combines the benefits of Mobile Cloud Computing and vehicular communications. In the VCC, drivers can access cloud services, e.g., processing services, from Service Providers (SPs). However, this may lead to disclosure of the drivers' personal information, such as location, identifiers, and routine, to the SPs or attackers. Therefore, the authors in~\cite{aloqaily2015auction} proposed a lightweight auction to provide cloud services which guarantee the low service cost, low service delay, and minimum amount of the information revealed to the SPs, while guaranteeing privacy preserving. The model consists of drivers as bidders, Trusted Third Parties (TTPs) as brokers, and SPs as sellers. Generally, the drivers submit the service requests along with the Quality of Experience (QoE) requirements to the TTPs. The QoE is expressed as a weighted function of delay, service cost, and privacy. Then, each TTP adopts the auction to select the best set of drivers to maximize the TTP's profit while meeting the drivers' QoE requirements. Similarly, each SP employs the auction model to select the best set of TTPs to maximize the SP's profit. Under the proposed solution, the profits of all parties will be maximized. In particular, the privacy preserving for the drivers is guaranteed through their QoE requirements


\subsection{Privacy leakage in spectrum allocation}
\label{sec:Privacy_preserving_spectrum_allocation} 
An efficient spectrum allocation scheme ensures that spectrum resource is assigned to the users which can use the resource most valuably. Auctions are among the most promising methods to achieve this objective since they always maintain the competition among users \cite{binmore2002biggest}, and further raising important sums of money for spectrum providers, i.e., the sellers. In spectrum auction, users act as bidders submit their bids to an auctioneer or a seller. Each bid contains information about a specific channel and the price that the user is willing to pay. Based on the bids, the auctioneer assigns the channels to the users and charges them according to which specific auction scheme is used. 

Note that in the spectrum auction, the same channel can be sold to multiple non-interfering users. As such, the users are required to provide their location information, in addition to their bids, to the auctioneer to construct the conflict constraints. This increases the location information leakage of users. In fact, the risk can also come indirectly from the bid submission of the users. For example, based on the channel specified in a bid of the user, an attacker can acquire a set of possible cells of the user by intersecting the complements of different seller's coverages \cite{liu2012adaptive}. Additionally, the price included in the bid can also increase the location leakage of the user. Indeed, bidding prices of the user for different channels typically imply the different channel quality estimations. The attacker can compute the difference between the estimated quality from the user and the real quality from a geo-location database in each candidate cell. The cell with minimum distance is regarded as the user's position. In summary, the location information and bids of users need to be protected in spectrum auctions. The following reviews privacy preserving approaches to address this issue. For convenience, they are classified according to the structure of the market that are the single-sided auction, i.e., one seller and multiple buyers, and the double-sided auction, i.e., multiple sellers and multiple buyers. 

\subsubsection{Single-sided auction}
\label{sec:Privacy_preserving_spectrum_allocation_single_auction} 
The pioneering work can be found in \cite{liu2013location} which is a two-stage privacy preserving scheme to address the location and bid leakage issue for Secondary Users (SUs) in the spectrum auction in Cognitive Radio Networks (CRNs). The model consists of a Trusted Third Party (TTP), an auctioneer as a spectrum owner, i.e., a seller, and SUs as bidders. 
The first stage focuses on the location and bid privacy preserving based on the TTP's secret keys which use the keyed-Hash Message Authentication Code (HMAC). The keys are known only by the SUs and the TTP. For the location privacy, each SU calculates its HMAC value based on its location and interference range. Then, the SUs submit the values to the auctioneer so that the auctioneer determines whether the SUs would interfere with each other. Similar to the bid privacy, the SU calculates its HMAC value for each channel bid, and then submits its HMAC value to the auctioneer. Based on the values, the auctioneer determines the winner for each channel using the prefix membership verification scheme \cite{chen2010safeq}. The auctioneer then assigns channels to the winners through a greedy spectrum allocation. In the second stage, the auctioneer delivers the masked winning bids encrypted by the symmetric key generated by the TTP and the corresponding prefix sets to the TTP. The TTP decrypts these bids to obtain their plaintext and then sends the plaintext back to the auctioneer so as to securely charge the winners. The simulation results showed that the proposed scheme reduces the location information leakage of the user from 20\% to 60\% compared to that in the case without using any privacy preserving. However, the use of the TTP increases the communication overhead which can affect real-time applications. Moreover, the proposed scheme did not consider the truthfulness of the auction. 

\begin{figure}[t!]
 \centering
\includegraphics[width=6.6cm, height= 5.4cm]{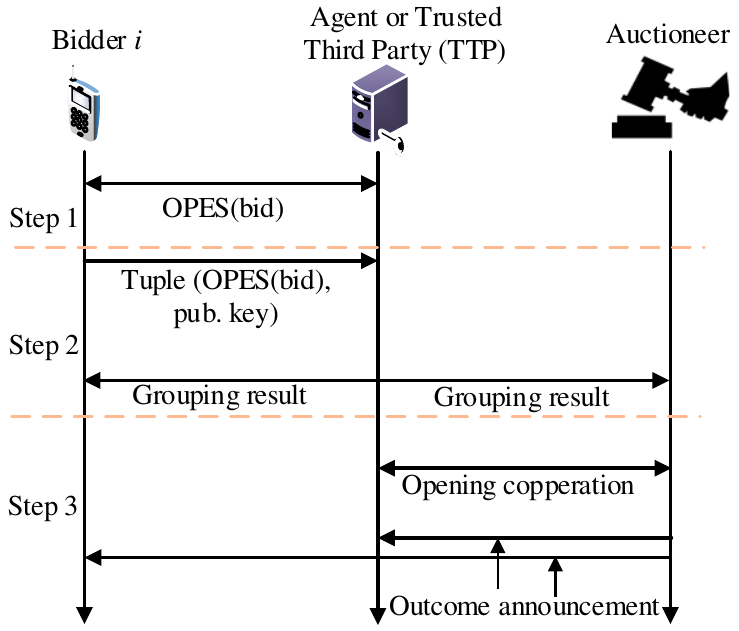}
 \caption{Bid privacy preserving with the help of the agent/TTP.}
 \label{privacy_concern_spectrum_allocation}
\end{figure}

The authors in \cite{wu2015towards} designed a spectrum auction by considering both the truthfulness and privacy preserving. The model is shown in Fig.~\ref{privacy_concern_spectrum_allocation} which consists of an auctioneer, i.e., the seller, a trustworthy authority which acts as an intermediate agent, and mobile users as bidders, i.e., buyers. The main idea is to separate the information known by the auctioneer and the agent such that no party has enough information to infer any sensitive information. The proposed auction is performed in three stages. In the first stage, bidders submit their bids based on their channel valuations to the agent. The agent uses the Order Preserving Encryption (OPE) \cite{agrawal2004order} to map the bids to another value while preserving the order of the bids. In the second stage, each bidder encrypts its value using the auctioneer's public key through an asymmetric encryption function. The bidder submits a tuple as an encrypted bid to the agent which includes its encrypted value and a signing function. The agent divides the bidders into non-conflicting groups, i.e., non-interfering groups, and publishes the grouping result and encrypted bids. In the third stage, the auctioneer decrypts the bids, calculates the prices representing for groups of bidders, and sorts the groups in a non-increasing order of their prices. The first $K$-bidder groups are the winners, and each winning bidder group is charged with a group bid of the $(K+1)$st group. This charging scheme enables the auction to achieve the truthfulness. The charge is then shared evenly among group members, i.e., bidders. 

The same approach can be found in \cite{huang2015general}. However, the Boneh-Goh-Nissim (BGN) cryptosystem involving the bilinear map and bilinear group \cite{boneh2005evaluating} was used instead of the OPE to map and encrypt the bids. The simulation results in \cite{huang2015general} showed that the satisfaction ratio, i.e., the percentage of winning bidders, of the proposed scheme can reach to one with the small number of bidders, i.e., less than 200. However, there is no numerical experiment to show the performance improvement related to the strategy-proofness as well as the privacy preserving. Moreover, the collusion between the agent and the auctioneer, as well as that among bidders, were also not mentioned. In fact, the 1-out-of-$n$ oblivious transfer protocol \cite{rabin2005exchange} can be used to encrypt the exchange messages between the bidders and the agent. This protocol prevents the agent from knowing the chosen bids. Otherwise, the secure multi-party computation \cite{prabhakaran2013secure} can be applied to allow the bidders in the same group to find the smallest bid for verifying the auction outcome while preserving the privacy of their bids.

By selecting the smallest bid as the representation bid of a non-conflict group, the approach in \cite{huang2015general} allows the bidders in the group to receive more benefit since they only pay less price than their expected bids. However, such a low price causes a loss in revenue for the seller. The authors in \cite{zhu2014differentially} combined the concept of the differential privacy with the spectrum auction in CRNs to preserve the privacy for bidders while obtaining approximately revenue maximization for the seller. The differential privacy aims to reveal information about the population as a whole, while protecting the privacy of each individual. The model involves several SUs as buyers, i.e., bidders, and one PU as the seller. The proposed scheme consists of three stages, i.e., \textit{grouping}, \textit{price determination}, and \textit{winner selection}. The grouping stage divides the bidders into non-conflicting groups using the graph coloring algorithms \cite{trudeau2013introduction}. Then, the price determination stage sets the probability of the final price of each group which is exponentially proportional to the seller's corresponding revenue. The revenue of each group is calculated based on the final price and the number of bidders in the group. The winner selection stage selects groups with the highest revenues as the winners and allocates them the channels.

It is worth noting that the pricing strategy in \cite{zhu2014differentially} can achieve good differential privacy for bidders' bids. This means that any change in a bidder's bid will not significantly change the revenue of the bidder's group, and thus the others cannot infer information of this particular bidder just from the revenues. In practice, a bidder can request more channels in the auction. In this scenario, the solution in \cite{huang2015general} can be applied. However, the bidder is first represented by elementary bidders, and each elementary bidder can be considered to be a buyer which requests one channel. The seller then divides the non-conflict groups based on these elementary bidders. Note that the elementary bidders belonging to the same bidder cannot use the same channel since they cause interference to each other. Finally, the solution in \cite{huang2015general} can be directly applied.

\begin{figure}[t!]
 \centering
\includegraphics[width=6.5cm, height= 5.4cm]{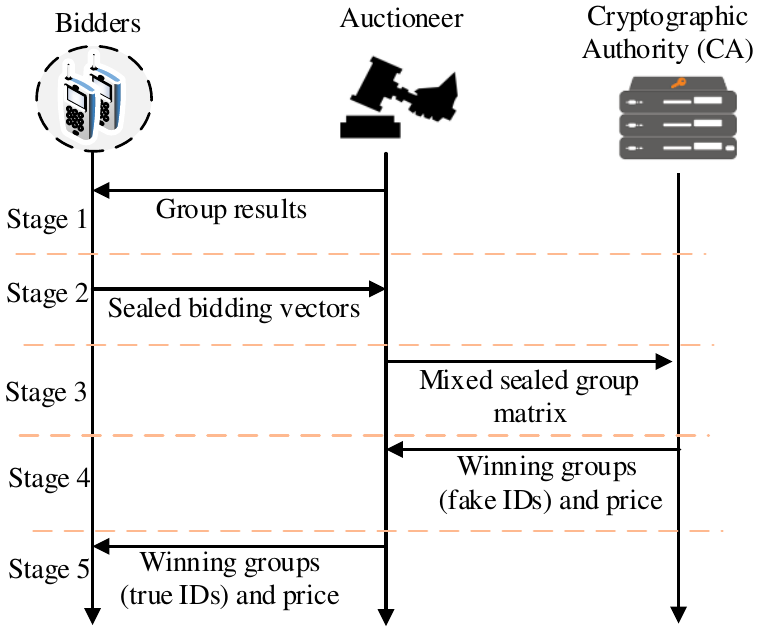}
 \caption{Privacy preserving protocol for bidders with the help of the CA.}
 \label{privacy_concern_cryptographic_authority}
\end{figure}


Apart from employing the OPE as proposed in \cite{wu2015towards}, the ElGamal encryption function \cite{elgamal1984public} can be used to preserve bid privacy for SUs in CRNs as proposed in \cite{wang2015privacy}. The ElGamal encryption function has the same homomorphic property as the Paillier cryptosystem mentioned in \cite{shi2013sealed}. The model consists of an auctioneer, a Cryptographic Authority (CA), a PU as the seller, and SUs as bidders. The proposed scheme has five stages as shown in Fig.~\ref{privacy_concern_cryptographic_authority} in which the first one is similar to the grouping stage in \cite{zhu2014differentially}. In the second stage, each bidder generates a \textit{sealed bidding vector} to conceal its actual bid using the homomorphic encryption generated by the CA. The bidders then submit their vectors to the auctioneer via a secure channel employing the auctioneer's public key. In the third stage, the auctioneer constructs a \textit{mixed sealed group matrix} from the sealed bidding vectors for each group using the homomorphic property. A random number is also introduced in the matrix to prevent the leakage of sensitive information. The auctioneer then sends all matrices of the groups to the CA with fake group identities. In the forth stage, the CA decrypts the minimal bid in each group by using the binary search tree algorithm \cite{cormen2009introduction} without requiring the other bid information in the group. Based on these minimal bids, the CA determines group bids, the winning groups, and their charges similar to those in \cite{wu2015towards}. Note that winning groups are charged by the same price, i.e., the uniform clearing price. In the fifth stage, the CA sends the set of winning groups with fake identities and the clearing price to the auctioneer. The auctioneer transforms the fake identities into the true ones and announces the winning groups along with the clearing price to all the bidders.  

It can be observed that in \cite{wang2015privacy}, since the homomorphic encryption key is generated by the CA, the auctioneer is unable to decrypt the sealed bidding vectors. On the other hand, the bidders send their sealed bidding vectors by using the public key of the auctioneer which will prevent the CA from eavesdropping. Therefore, the proposed scheme guarantees that no sensitive information other than group bids is exposed as long as the auctioneer and CA do not collude with each other. However, by generating the ciphertext for every possible price, the communication overhead of bidders is extremely high. As shown in the simulation results, when the number of possible prices equals 1000, the communication overhead of the proposed approach is around 62.5KB while that of the no-privacy-preserving spectrum auction scheme \cite{zhou2009trust} is only 0.001KB. 

\subsubsection{Double-sided auction}
\label{sec:Privacy_preserving_spectrum_allocation_double_auction} 
To protect the users' location information, another possible technique is the $k$-anonymity location privacy \cite{gruteser2003anonymous}. With the $k$-anonymity location privacy, user's location information is $k$-anonymous, meaning that the user cannot be distinguished from the location information of at least $(k-1)$ other users. To provide $k$-anonymity location privacy, it is required to have an anonymity set consisting of at least $k$ users. However, some users may not be sensitive about their own location information, and they may have no incentive to participate in the anonymity set. When there are not enough privacy-sensitive users in the anonymity set, the privacy-sensitive users can act as buyers and invite users which are not sensitive to privacy, i.e., sellers, to participate. 

To provide the sellers incentives to take part in the anonymity set while achieving some economic properties such as the balanced budget, truthfulness, and high satisfaction ratio, the authors in \cite{zhang2016designing} adopted the double auction to model the interaction between the buyers, the sellers, and an auctioneer, i.e., a central authority. The number of buyers is assumed to be less than $k$ and to have the same privacy degree requirements. The proposed approach consists of two stages. In the first stage, the buyers submit to the auctioneer their bids, i.e., bidding prices that they are willing to pay. Simultaneously, the sellers submit to the auctioneer their asks, i.e., asking prices that they agree to participate in the anonymity. The auctioneer then sorts the buyers in a descending order of bids and the sellers in an ascending order of asks. The auctioneer finds the largest index $x$ at which the sum of bids of $x$ buyers is not less than the sum of asks of $(k-x)$ sellers. This is to satisfy the balanced budget property of the auction. The $x$ first buyers and the $(k-x)$ first sellers are the winners. In the second stage, the charges for the winning buyers and the payments for the winning sellers are determined according to the ask of $(k-x+1)$th seller. This pricing strategy guarantees the truthfulness of the proposed scheme. The simulation results also showed that the proposed scheme outperforms the baseline in \cite{yang2013truthful} in terms of the satisfaction ratio, i.e., the percentage of buyers winning the auction. However, the auctioneer's profit in the proposed approach is zero or negative in some cases. 

The truthful double auction can also be found in \cite{chen2014ps} which addressed the privacy preserving for users in the spectrum allocation. In this model, the users which act as buyers, i.e., bidders, buy channels from the other users, i.e., sellers, via an auctioneer and an auction agent. The proposed algorithm includes three steps. The first step forms non-conflict groups of the bidders as implemented in \cite{wu2015towards}. The agent publishes the public key based on the Paillier cryptosystem to the auctioneer and the bidders. The bidders use the public key to map their bids to Encrypted Bit Vectors (EBVs). In the second step, the bidders submit the EBVs to the auctioneer. The auctioneer uses the homomorphic property of the Paillier cryptosystem to find the minimum EBV bid and compute the EBV group bid without the knowledge of the bidders' bids. Then, the auctioneer finds (i) the encrypted bidder group index with the highest group bid, and (ii) the seller index with the lowest ask. The two indexes are sent to the agent which performs the decryption to obtain actual indexes. The agent sends the decrypted information back to the auctioneer. If the highest group bid is not less than the lowest ask, the corresponding bidder group and the seller are the winners. This process is repeated for the remaining bidder groups and sellers. The group bid and the ask obtained in the last iteration of the process are the buying and selling clearing prices, respectively. In the third step, each bidder in the winning bidder group pays the equal share of the buying clearing price. 

The proposed algorithm in \cite{chen2014ps} achieves the correctness, meaning that the double auction result is the same as the result of double auctions without the cryptosystem. However, it has three shortcomings. First, it does not consider the privacy preservation of sellers' asks. Second, the ranking orders of the winning sellers' asks and the winning buyer group's bids are leaked to both the auctioneer and the agent, i.e., the proposed algorithm is not really secure. Third, the efficiency of the proposed algorithm is not satisfactory in terms of time complexity. As shown in the simulations, when the number of sellers is 30, and that of buyers is 70, the total time to receive the final result is 80 minutes, which is unacceptable. 

The first shortcoming can be simply solved by applying the Paillier cryptosystem to the asks as proposed in \cite{chen2017secure}. The second shortcoming can be addressed by hiding the ranking orders of winners. One possible solution is that the auctioneer uses the set of randomized sellers and the set of randomized buyers instead of the original ones. Thus when the agent sends actual indexes of a bidder group-seller pair to the auctioneer, the auctioneer does not know which pair is. For the third shortcoming, the Batcher's sorting network \cite{knuth1998art} can be used to limit the number of bidder groups and sellers before matching each bidder group's bid and each seller's ask as proposed in \cite{li2015}. 

\subsection{Bid Integrity for Mobile Ad Hoc Networks (MANETs)}
\label{sec:Survey_Con_In_Au_Mess_Inte}
The bid privacy preserving schemes as discussed in Sections \ref{sec:Privacy_preserving_data_collection} and \ref{sec:Privacy_preserving_spectrum_allocation} are always performed in networks in which central
management entities such as the TTP and the auctioneer are easily deployed. However, in infrastructure-less networks such as MANET, the schemes are almost infeasible due to the absence of central management entities. Thus bids are easily revealed, modified, or damaged during the bidding process. Indeed, to develop an auction process in the MANET, a bidder can submit its bid directly to a seller. If the bidder cannot communicate directly with the seller, it asks another bidder as a router to forward its bid to the seller. When forwarding the bid, a malicious bidder can modify the bid of a rival bidder if the bid is not protected. Therefore, the bid integrity which ensures that the bidder's bid is not modified or damaged during transmission needs to be considered. 

\begin{figure}[t!]
 \centering
\includegraphics[width=4.9cm, height = 6.1cm]{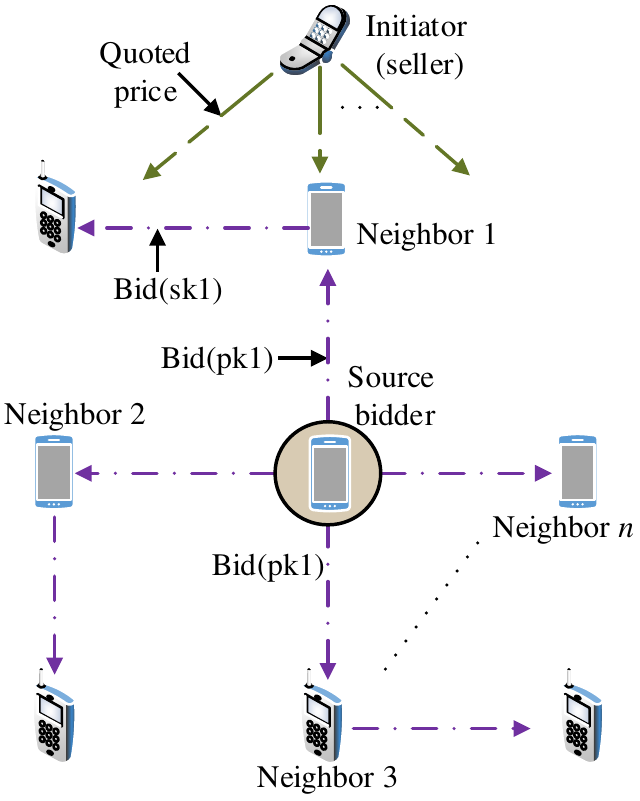}
 \caption{Bid's integrity in MANET based on the $(k,n)$-threshold cryptography scheme. Bid(pk1) is the source bidder's bid encrypted with neighbor 1's public key, and bid(sk1) is the source bidder's bid signed with neighbor 1's secret key.}
 \label{bid_integrity_english_auction}
\end{figure}

The authors in \cite{fourati2006deploying} investigated the bid integrity issue for the English auction in MANET. The model is shown in Fig.~\ref{bid_integrity_english_auction} in which the initiator acts as a seller which trades a resource unit, and network nodes, i.e., bidders, compete on the resource. In the English auction, the seller initially sets the lowest price of the resources and increases the price in the next rounds. The auction terminates if there is no new higher price submitted, and the bidder with the highest bid wins the resource. To guarantee the integrity of the bids, the Shamir's $(k,n)$-threshold cryptography scheme \cite{shamir1979share} was used. Accordingly, a private key is shared among $n$ bidders in such a way that no single bidder can recover the private key without the cooperation of $(k-1)$ other bidders. One example is shown in Fig.~\ref{bid_integrity_english_auction}. The source bidder first shares its private key with $n$ one-hop neighbors, and then sends its bid to these neighbors in unicast messages encrypted with the neighbors' public keys. Each neighbor signs the received bid with its own secret key which is generated based on the source bidder's private key. The neighbors then flood the signed bids to all bidders in the network. Given the property of the $(k,n)$-threshold cryptography, a bidder which wants to recover the original bid of the source bidder must receive at least $k$ versions of the bid with $k$ different secret keys.


Apart from guaranteeing the bid integrity, the bidding process in \cite{fourati2006deploying} needs to prevent a network attack which commonly occurs in the bid forwarding, called \textit{replay} or \textit{playback} attack. In the replay attack, a malicious bidder intercepts the bids from the others and submits them again in a future auction round \cite{lin2015vehicular}. Thus the malicious bidder may be considered to be a legitimate bidder, but then it disturbs the routing/forwarding operation in the network. To prevent the replay attack, the authors in \cite{ayed2012fairness} added a so-called \textit{nonce} in each bid. In cryptography, a nonce is an arbitrary number which may only be used once. The nonce in each bid is generally calculated based on the identifier of the current auction round and the private key of the corresponding bidder. The nonce's value is unique in each auction round and invalid in other auction rounds. Thus a malicious bidder cannot reuse another bidder's bid in a next auction round. However, this solution requires the initiator and each bidder to create and manage an Access Control List (ACL) which contains the identifiers of all bidders in the network. The ACL also needs to be updated when a bidder joins or leaves the network. This significantly increases communication overhead. 

\textbf{Summary:} This section discusses privacy preserving mechanisms for users through integrating encryption methods with pricing models. The reviewed approaches along with their references are summarized in Table~\ref{table_privacy_concern}. We observe that almost all the approaches investigate the integration of auction schemes with cryptographic algorithms to protect the information privacy for bidders. Also, the spectrum allocation receives more attentions than the sensing data collection. In fact, the spectrum allocation also faces other security issues which severely
deteriorate the efficiency of the allocation. The security issues commonly arise from the cheating
and illegitimate behaviors of users in the network, i.e., the collusion and the false-name bid cheating. The following section reviews the pricing approaches to deal with these issues.

\begin{table*}
\caption{Applications of economic and pricing models for privacy concerns and confidentiality (CN: Cellular Network, MANET: Mobile Ad hoc NETwork,  MCN: Mobile Crowdsensing Network)}
\label{table_privacy_concern}
\scriptsize
\begin{centering}
\begin{tabular}{|>{\centering\arraybackslash}m{0.2cm}|>{\centering\arraybackslash}m{0.4cm}|>{\centering\arraybackslash}m{1.4cm}|>{\centering\arraybackslash}m{1cm}|>{\centering\arraybackslash}m{1cm}|>{\centering\arraybackslash}m{1cm}|>{\centering\arraybackslash}m{5.2cm}|>{\centering\arraybackslash}m{2.5cm}|>{\centering\arraybackslash}m{1.2cm}|>{\centering\arraybackslash}m{0.8cm}|}
\hline
\multirow{2}{*} {\textbf{}} & \multirow{2}{*} {\textbf{Ref.}} & \multirow{2}{*} {\textbf{Pricing model}} & \multicolumn{3}{c|} {\textbf{Market structure}} & \multirow{2}{*} {\textbf{Mechanism}} & \multirow{2}{*} {\textbf{Objective}} & \multirow{2}{*} {\textbf{Solution}}& \multirow{2}{*} {\textbf{Network}} \tabularnewline
\cline{4-6}
 & & & \textbf{Seller} & \textbf{Buyer} & \textbf{Item} & & &&\tabularnewline
\hline
\hline
\parbox[t]{2mm}{\multirow{9}{*}{\rotatebox[origin=c]{90}{\hspace{-4cm} Privacy concerns in sensing data collection}}}

& \cite{singla2013incentives}  &Reverse Vickrey auction &Phone users &Platform & Sensing data&Sellers submit their asking prices with obscure locations. The buyer selects the winners with the lowest asking prices.&Incentive
compatibility, revenue improvement, and privacy preserving &Nash equilibrium&MCN\tabularnewline \cline{2-10}

& \cite{sun2014privacy}  &Reverse Vickrey auction &Phone users &Server & Real identity& Sellers' asks are encrypted by the TLC method and signed by using the Nyberg-Rueppel signature scheme. The sellers with asks lower than a threshold are selected as the winners. &truthfulness, revenue improvement, and privacy preserving &Nash equilibrium&MCN\tabularnewline \cline{2-10}

& \cite{wang2016incentive}  &General auction &Phone users &Server & Sensing data& Same as \cite{sun2014privacy}, but the winner selection is based on the past reputation and marginal utility of sellers.&truthfulness, efficiency, and privacy preserving&Nash equilibrium&MCN\tabularnewline \cline{2-10}

& \cite{wu2016magicrowd}  &General auction &Phone users &Campaigner & Sensing data& The auctioneer clusters the sellers into crowds and bargains with the buyer on the deal price for each crowd using the bargaining game.&truthfulness, anonymity, and fairness &Nash bargaining equilibrium&MCN\tabularnewline \cline{2-10}  

& \cite{shi2013sealed} &Multi-attribute reverse auction &Phone users &Server &Sensing data& Each ask is encrypted by Paillier cryptosystem. The buyer determines the winner by using the private set intersection algorithm. &Anonymity, public verifiability, and fairness &Bayes-Nash equilibrium&MCN\tabularnewline \cline{2-10}

& \cite{dimitriou2015privacy} &Multi-attribute reverse auction &Phone users &Server &Sensing data& Each ask with multiple attributes corresponding to a utility score is masked with a hash value. Sellers with the highest scores are the winners. &Privacy preserving, public verifiability, and fairness &Bayes-Nash equilibrium &MCN\tabularnewline \cline{2-10}

&\cite{holzbauer2012socially}&First-price sealed-bid reverse auction &Phone users &Data sink &Sensing data& Each seller's ask is generated based on its power level and a privacy function. The sellers with the lowest asks are the winners. &Privacy preserving, and energy balance &Nash equilibrium &MCN\tabularnewline \cline{2-10}

\hline
\parbox[t]{2mm}{\multirow{9}{*}{\rotatebox[origin=c]{90}{\hspace{-7cm} Privacy concerns in spectrum allocation}}}

& \cite{liu2013location}&First-price sealed-bid auction &Spectrum owner&SUs &Spectrum& Buyers' bids are encrypted by the HMAC values of the TTP. The auctioneer uses the prefix membership verification to determine a winner. Then, the auctioneer cooperates with the TTP to securely charging the winner. &Privacy leakage reduction, and satisfaction ratio improvement &Nash equilibrium& CRN\tabularnewline \cline{2-10}

&\cite{wu2015towards}&Vickrey auction &Spectrum supplier &Phone users &Spectrum& Buyers' bids are encrypted by the agent's OPE. The auctioneer cooperates with the agent to determine the winners and charges them according to the second-price auction rule. &Privacy preserving, satisfaction ratio improvement, and truthfulness &Nash equilibrium& CN\tabularnewline \cline{2-10}

&\cite{huang2015general}&Vickrey auction&Spectrum supplier &Phone users &Spectrum&Same as \cite{wu2015towards}, but the BGN cryptosystem is used to encrypt the buyers' bids. &Privacy preserving, satisfaction ratio improvement, and truthfulness &Nash equilibrium&CN\tabularnewline \cline{2-10}

&\cite{zhu2014differentially}&General sealed-bid auction &PU &SUs &Spectrum&Seller applies the differential privacy to set the probability of the final price of each buyer group according to the seller's revenue. & Differential privacy, and seller's revenue maximization&Nash equilibrium&CRN\tabularnewline \cline{2-10}

&\cite{wang2015privacy}&Vickrey auction &PU &SUs &Spectrum&Buyers' bids are encrypted by the ElGamal encryption function. The auctioneer constructs a mixed sealed group matrix for each bid. The auctioneer cooperates with the CA to determine winning groups and their charges. & Privacy preserving, truthfulness, and spectrum utilization efficiency&Nash equilibrium&CRN\tabularnewline \cline{2-10}

&\cite{zhang2016designing}&Double auction &Phone users &Phone users &Participation&$k$ winning buyers and winning sellers are determined to satisfy the balanced budget. The charge and the payment are identically set to satisfy the truthfulness.& $k$-anonymous location privacy, truthfulness, balanced budget, and satisfaction ratio improvement&Market equilibrium&CN\tabularnewline \cline{2-10}

&\cite{chen2014ps}&Double auction &Phone users &Phone users &Spectrum&Buyers' bids are encrypted using the Paillier cryptosystem. The auctioneer cooperates with the agent to determine winning bidder groups, i.e., winning bidders, winning sellers, and clearing price. & Privacy preserving, correctness, and truthfulness&Market equilibrium& CN\tabularnewline \cline{2-10}

&\cite{li2015}&Double auction &Phone users &Phone users &Spectrum&Same as \cite{chen2014ps}, but the Batcher's sorting network is used to limit the number of asks and bids before the winner determination process.& Privacy preserving, correctness, and efficiency&Market equilibrium&CN\tabularnewline \cline{2-10}

\hline
\parbox[t]{2mm}{\multirow{9}{*}{\rotatebox[origin=c]{90}{\hspace{1cm} Bid integrity}}}

& \cite{fourati2006deploying}&English auction &Initiator&Network nodes&Network resource& Steps are similar to those in an English auction, but each buyer's bid is encrypted by the $(k,n)$-threshold cryptography. &Revenue improvement, and bid integrity &Nash equilibrium&MANET\tabularnewline \cline{2-10}

&\cite{ayed2012fairness}&English auction &Initiator&Network nodes&Network resource& Same as \cite{fourati2006deploying}, but the identifier of the current auction round and the private key of each buyer are included in buyer's bid.  &Revenue improvement, bid integrity, fairness, and replay attack prevention&Nash equilibrium &MANET\tabularnewline \cline{2-10}

\hline
\end{tabular}
\par\end{centering}
\end{table*}

\begin{table*}[h]
\caption{A summary of advantages and disadvantages of major approaches for privacy concerns and confidentiality}
\label{table_sum_advantage_privacy}
\scriptsize
\begin{centering}
\begin{tabular}{|>{\centering\arraybackslash}m{1.8cm}|>{\centering\arraybackslash}m{6.5cm}|>{\centering\arraybackslash}m{8.3cm}|}
\hline
\cellcolor{myblue} &\cellcolor{myblue} &\cellcolor{myblue} \tabularnewline
\cellcolor{myblue} \multirow{-2}{*} {\textbf{Major approaches}} &\cellcolor{myblue} \multirow{-2}{*} {\textbf{Advantages}} &\cellcolor{myblue} \multirow{-2}{*}{\textbf{Disadvantages}} \tabularnewline
\hline
\hline
\cite{liu2013location}&\begin{itemize} \item Preserve privacy for both location and bid price  \end{itemize} & \begin{itemize}  \item Have high communication overhead \end{itemize}\tabularnewline \cline{2-3}
  \hline
\cite{wu2015towards} &  \begin{itemize} \item Achieve both privacy preserving and truthfulness \end{itemize}& \begin{itemize} \item Ignore the collusion between the auctioneer and the trustworthy authority \item Do not consider the loss in the seller's revenue  \end{itemize}\tabularnewline \cline{2-3}
  \hline
\cite{zhu2014differentially} & \begin{itemize} \item Achieve privacy preserving and revenue maximization \end{itemize} & \begin{itemize} \item Support reserving only a single channel for a buyer \end{itemize} \tabularnewline \cline{2-3}
  \hline
\cite{wang2015privacy} &\begin{itemize} \item Achieve both privacy preserving and truthfulness\end{itemize}&\begin{itemize} \item  Ignore the collusion between the auctioneer and the cryptographic authority  \item Have high overhead communication  \end{itemize} \tabularnewline \cline{2-3}  
  \hline
\cite{chen2014ps}& \begin{itemize}\item Support preserving privacy for multiple phone users  \item Achieve privacy preserving and correctness  \end{itemize}& \begin{itemize} \item Have high time complexity and do not consider the ask privacy  \end{itemize} \tabularnewline \cline{2-3}
  \hline
 \cite{li2015}& \begin{itemize}\item Support preserving privacy for multiple phone users  \item Have low time complexity  \end{itemize}& \begin{itemize} \item Ignore the collusion among bidders  \end{itemize} \tabularnewline \cline{2-3}
  \hline
\cite{fourati2006deploying}& \begin{itemize}\item Be simple to be implemented  \end{itemize}& \begin{itemize} \item Do not guarantee the fairness among bidders \end{itemize} \tabularnewline \cline{2-3}
\hline
\end{tabular}
\par\end{centering}
\end{table*}

\section{Illegitimate behaviors in wireless networks}
\label{sec:Survey_collusion_false}
This section reviews applications of pricing models to prevent the following illegitimate behaviors in spectrum auctions.
 \begin{itemize}
\item \textit{Collusion:} In spectrum auctions, the bidders may collude with each other through coordinating their bids which suppresses the competition for spectrum resources. The low competition degrades the spectrum efficiency as well as the seller's revenue. To prevent the collusion, the system approach, e.g., \cite{wang2013one}, can be applied. However the system approach is considered to be the "hard" approach that requires additional hardware and software to implement and enforce it. This can incur substantial cost and overhead. More importantly, since the users still have interest to collude, they can adapt and find any flaw in the system approach to successfully make a collusion. Conversely, the economic approach is considered to be the "soft" approach that motivates and incentivizes the users not to make collusion as it is not profitable for them to do so. Therefore, the economic approach is natural and inherent in the decision making of the users, thus more efficient and more effective to employ.
\item \textit{False-name bids:} A single bidder can increase its utility by submitting false bids made under multiple fictitious identifiers. This may lead to a severe loss in revenue for the seller. Pricing models are adopted to guarantee that a bidder cannot increase its utility with false-name bids.
\end{itemize}

\subsection{User Collusion and Bid-rigging}
\label{sec:Survey_collusion_false_coll}
\begin{figure}[t!]
 \centering
\includegraphics[width=6.5cm, height=4.5cm]{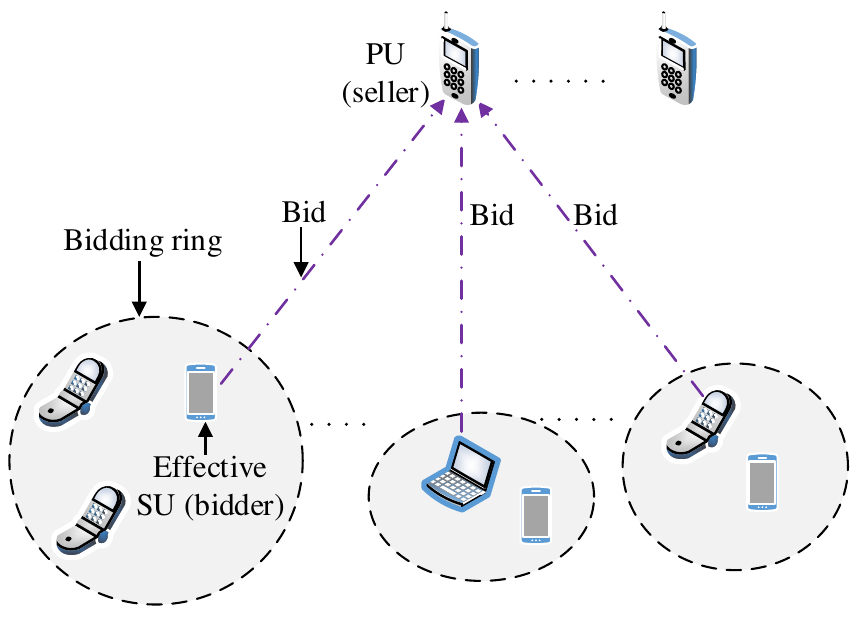}
 \caption{Collusion-resistance based on the second-price sealed-bid auction.}
 \label{collusion_second_price}
\end{figure}

The authors in \cite{ji2007collusion} adopted the Vickrey auction with the aim of improving the utility of a Primary User (PU), i.e., the seller, in the presence of the collusion among Secondary Users (SUs), i.e., bidders, as shown in Fig.~\ref{collusion_second_price}. First, the SUs are divided into bidding rings, each of which is a collection of bidders which collude in the auction. In each bidding ring, the SU with the highest bid, called the effective SU, represents the bidding ring. The PU selects the effective SU with the highest bid as the winner. The reserve price for the winner is determined to maximize the expected utility gain of the PU. In particular, the gain is formulated based on the Cumulative Distribution Functions (CDFs) of the highest and second highest bids of the effective SUs. Such pricing strategy guarantees that the PU always receives a price greater than or equal to the optimal reserve price, and thus its revenue is high. The simulation results showed that given the same collusion rate, the proposed algorithm outperforms the one without using the reserve price in terms of the PU's utility. However, how to determine the bidding rings as well as the statistics, i.e., the CDFs, of bids was not explained. 

In fact, the solution in \cite{ji2007collusion} can also be applied to the scenario involving multiple PUs and one SU. In this case, the PUs, i.e., the sellers, collude with each other to obtain higher prices. Similar to the scenario with multiple SUs and one PU, the PUs are first divided into the bidding rings. In each bidding ring, the PU with the lowest price is selected as the effective PU which represents for the bidding ring. The SU formulates its expected utility gain with the reserve price which it is willing to pay for leasing the channel. The optimal reserve price is then determined by taking the first-order derivative of maximizing the expected utility gain.

Considering a more general scenario with multiple
PUs and multiple SUs, the authors in \cite{ji2008multi} adopted the double auction to combat the collusion among the SUs and the collusion among the PUs. This case can be considered to be the combination of the two scenarios discussed in \cite{ji2007collusion}. Thus the optimal reserve price for each PU or SU can be determined using the results in \cite{ji2007collusion}. Additionally, to obtain the CDFs of the effective SUs' bids and the effective PUs' asks, the common belief functions at different prices of the PUs and the SUs are constructed. The common belief function of the PUs/SUs at a certain price is defined as the ratio of asks/bids from the PUs/SUs at the price that has been accepted. The optimal reserve prices for each PU and SU are calculated based on the CDFs. Also, the PUs and SUs determine their asks and bids to maximize their expected payoffs, respectively. If a bid of an SU is greater than or equal to an ask of a PU, the channel leasing agreement is performed between them. The process continues until the spectrum pool of the PUs is empty. 

The spectrum auction proposed in \cite{ji2007collusion} guarantees that the PU's revenue is not too low. However, this approach is used only for the PU with a single channel. The PU may have more than one channel, and the multi-winner auctions such as the VCG auction can be applied. However, the VCG auction aims to maximize the social welfare rather than to increase the PU's revenue. Moreover, the VCG auction is susceptible to the collusion \cite{bachrach2010collusion}. One reason is that the prices set by the auction are sometimes too low, and the losers can afford the prices. For example, the losers can collude to win channels and have sufficient margins to make extra profits by subleasing the channels to others. 

 To prevent the collusion among SUs while improving the PU's revenue and maximizing the spectrum utilization, the authors in \cite{wu2008multi} remodeled the multi-winner spectrum auction into a single-winner spectrum auction by introducing the concept of \textit{virtual bidders}. Specifically, SUs without mutual interference are grouped into a virtual bidder, the bid of which equals the sum of the individual bids. The virtual bidder with the highest bid is the winner of the channel. The winner pays the PU the price equal to the bid of the highest losing virtual bidder. Then, the payment among SUs within the winning virtual bidder is calculated so as to maximize the product of the individual payoffs. This ensures that the profits are shared among the SUs as equally as possible, and the PU's revenue is relatively high. Moreover, if some losers collude to beat the winners by raising their bids, the losers will have to pay a high price. Thus the loser collusion is eliminated. The simulation results showed that with the proposed pricing strategy, the colluding gains drop significantly compared with that based on the VCG auction.

Although the solution in \cite{wu2008multi} avoids the collusion among the losers, there may be the \textit{sublease collusion} among the winning SUs and the losing SUs. For example, the winning SUs sublease their channels to the losing SUs at acceptable prices to gain their revenue. Such collusion deteriorates the spectrum efficiency and the PU's revenue. However, the sublease collusion may be avoided if the prices to the winning SUs are set such that the total payment of the winning SUs is larger than the total payment of the losing SUs. The authors in \cite{wu2009scalable} imposed this condition instead of using the constraint of the optimization problem in \cite{wu2008multi}. The optimization problem now becomes convex and is efficiently solved by the numerical methods \cite{boyd2004convex}. In practice, the collusion among the losers happens more often than that among the winners since the payoffs of the losers are typically not fulfilled. Thus the auctioneer may provide the losers more incentives, but this may make the auctioneer's budget not balanced.

\begin{figure}[t!]
 \centering
\includegraphics[width=5.6cm, height= 5.4cm]{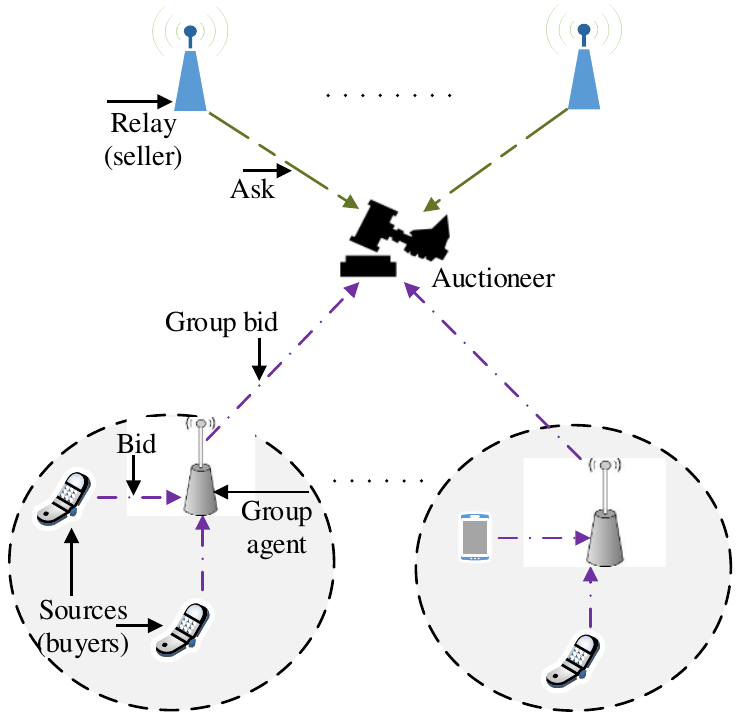}
 \caption{Collusion-resistance based on double auction.}
 \label{collusion_double_auction}
\end{figure}

To guarantee both the budget-balance and the collusion-resistance, the authors in \cite{xu2014collusion} adopted the double auction when assigning relay nodes to sources in cooperative wireless networks. The model is shown in Fig.~\ref{collusion_double_auction} where the sources act as buyers, and the relays are sellers. The sources which can share a relay are merged into a group and represented by a so-called \textit{group agent}. First, each source submits a set of its bids for all relays to its group agent. The group agent determines the candidate winners and the losers within its group. The agent also recalculates group bids for relays and incentives reserved for the losers. The incentives for the losers are extracted proportionally to the bids of the candidate winners. Then, the group agents and the relays submit theirs group bids and asks to the auctioneer, respectively. The auctioneer determines the winning group agents and the winning relays based on the maximum weight matching \cite{yang2011opra}. Since the losing group agents can also collude with each other, the incentives for them are required to be proportional to the auctioneer's revenue. 
 
Apart from the collusion among the bidders, there is another form of collusion in which the cheating auctioneer conspires with greedy bidders to illegally fix the price, namely the \textit{bid-rigging}. To prevent bid-riggings in spectrum allocation, the authors in \cite{pan2012using} adopted a variant of VCG auction integrated with the homomorphic encryption. Specifically, each bidder, i.e., an unlicensed user, submits its identity, location information, and bid for a set of spectrum bands to the auctioneer, i.e., the seller. In particular, the bid is represented by a vector of ciphertexts generated by the homomorphic encryption. The auctioneer determines the winners based on the encrypted bids using the homomorphic addition property. Based on the VCG auction, the auctioneer then determines the price of each set of spectrum bands for each bidder using the Shamir's polynomial secret sharing \cite{suzuki2002secure}. Given the prices, each bidder decides a set of spectrum bands to maximize its own utility. The security analysis stated that unless the servers which are deployed to encrypt/decrypt bids collude, the auctioneer cannot decrypt vectors representing the bidders' bids. Also, the bid-rigging between the bidders and the auctioneer is meaningless since the individual server itself knows nothing more than the winners' identities and their payments. 

The same approach can be found in \cite{abdelhadi2015multitier} in which the auctioneer is a broker, i.e., the seller, and the greedy bidders are Base Stations (BSs). In this approach, the Paillier cryptosystem was used to generate vectors of ciphertexts for the bidders' bids. Moreover, the encrypted bids are randomized by a random constant which is generated by an intermediate federal gateway. From these randomized encrypted bids, the auctioneer determines the winners and selects an optimal spectrum allocation so as to maximize the sum of the bids. Given the optimal allocation, the charging price for each winner is determined according to the VCG auction-based pricing policy. Note that with the homomorphic addition property of the Paillier cryptosystem, the auctioneer determines the winning allocation and the charging prices without any knowledge of the original bids of the BSs. Thus the bid-rigging between the auctioneer and a greedy bidder is avoided. In fact, this can be guaranteed even if the bidder colludes with the auctioneer since its bid is randomized by the federal gateway.

\subsection{False-Name Bids}
\label{sec:Survey_collusion_false_false}
The key idea to prevent the false-name bid cheating of a bidder is to make the utility of the bidder not increase by using multiple fictitious identifiers compared to the case in which the bidder uses a single identifier. 

\begin{figure}[t!]
 \centering
\includegraphics[width=\linewidth]{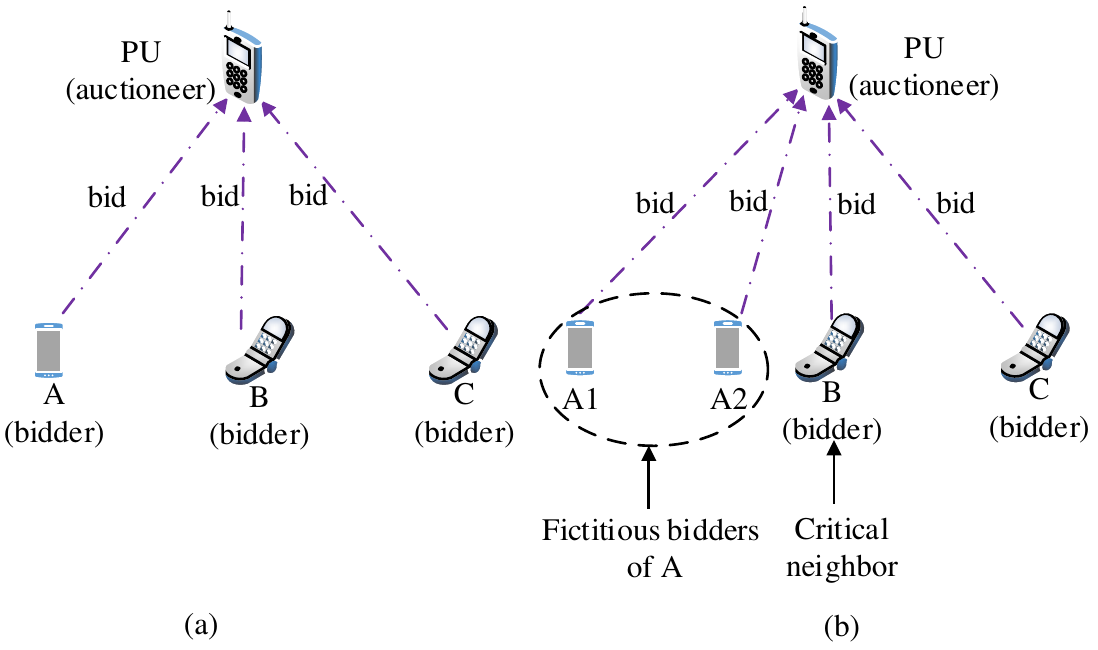}
 \caption{An example of spectrum auction without false-name bid cheating (a), and with false-name bid cheating (b). When bidder A uses two names, i.e., A1 and A2, to bid for the same number of channels, its utility increases.}
 \label{false_name_bid_cheating}
\end{figure}

The authors in \cite{wangrobust} proposed a pricing strategy for each SU in CRNs to prevent the false-name bid cheating in the spectrum auctions as shown in Fig.~\ref{false_name_bid_cheating}. The proposed approach consists of three steps. In the first step, each SU submits its bid for a set of channels to an auctioneer, i.e., a PU. To maximize the spectrum reusability while satisfying the interference constraints, the breadth-first-search procedure \cite{kurant2010bias} was applied to sort the bidders in a descending order of their bids. In the second step, the PU computes the price for each bidder by multiplying the bidder's bid with the bid of its \textit{critical neighbor}. A critical neighbor of a bidder is one of its interfering/conflicting neighbors that if the bidder bids higher than that of the critical neighbor, the bidder will receive a channel. However, if the bidder bids lower than the critical neighbor, the bidder will not be allocated a channel. In the third step, the PU compares the bid of each bidder with its computed price. If the bid is greater than its computed price, the bidder is the winner. Otherwise, the bidder is the loser.

The pricing strategy in \cite{wangrobust} was proved as to guarantee that a bidder using a single identifier obtains its requested channel with a utility that is not less than the sum of utilities obtained by using more than one identifier. The reason is that although the bidder has multiple identifiers, these identifiers come from the same location as the single identifier. Therefore, they have the same critical neighbor and the same computed price. The simulation results showed that the proposed scheme significantly improves revenue and spectrum utilization up to 300\% compared to the baseline \cite{terada2003false} which sacrifices spectrum reuse to maintain the false-name-proofness. However, the proposed scheme did not support the privacy preserving for bids. 

To support the privacy preserving while avoiding the false-name bids, the authors in \cite{rathinakumar2016gavel} adopted the ascending-bid auction which is essentially similar to the ACA scheme in \cite{zhang2013ascending}. The model includes a Licensed Shared Access (LSA) repository, i.e., an auctioneer, and LSA licensees, i.e., bidders. Assume that the conflict graph was created. Initially, the auctioneer announces a reserve price vector for a vector of channels. Each bidder responds with a demand vector which includes portions of channels. At each round, for a channel, the auctioneer determines if for any bidder the aggregate demand of the bidder and its neighbors in the conflict graph is less than the channel supply. If so, the portion of the channel is assigned to the bidder. This process repeats by increasing the round prices of channels until there is no demand for the channels from the bidders. It is worth noting that a bidder and the corresponding fictitious bidder have the same set of neighbors. Thus unless the total demand from the neighbors changes, a bidder cannot increase its utility with false-name bids. Also, since the bidders only reveal their demand but not prices, the proposed scheme achieves the price privacy for the bidders.

\textbf{Summary:} This section discusses applications of pricing models to prevent the collusion and false-name bidding cheating. The reviewed approaches along with their references are summarized in Table~\ref{table_collusion_false_bid_name}. As seen, almost all the approaches aim to prevent the collusion among the bidders or between the bidders and the auctioneer in the spectrum allocation. Therefore, more solutions to prevent the collusion behaviors of sellers need to be considered. In the next section, we review applications of economic and pricing models to address some other security issues in wireless networks.

\begin{table*}
\caption{Applications of economic and pricing models for protecting against collusion and false-bid name cheating behaviors (CN: Cellular Network, CRN: Cognitive Radio Network, CWN: Cooperative Wireless Network)}
\label{table_collusion_false_bid_name}
\scriptsize
\begin{centering}
\begin{tabular}{|>{\centering\arraybackslash}m{0.2cm}|>{\centering\arraybackslash}m{0.4cm}|>{\centering\arraybackslash}m{1.3cm}|>{\centering\arraybackslash}m{0.8cm}|>{\centering\arraybackslash}m{1cm}|>{\centering\arraybackslash}m{0.8cm}|>{\centering\arraybackslash}m{5.8cm}|>{\centering\arraybackslash}m{2.6cm}|>{\centering\arraybackslash}m{1.2cm}|>{\centering\arraybackslash}m{0.8cm}|}
\hline
\multirow{2}{*} {\textbf{}} & \multirow{2}{*} {\textbf{Ref.}} & \multirow{2}{*} {\textbf{Pricing model}} & \multicolumn{3}{c|} {\textbf{Market structure}} & \multirow{2}{*} {\textbf{Mechanism}} & \multirow{2}{*} {\textbf{Objective}} & \multirow{2}{*} {\textbf{Solution}}& \multirow{2}{*} {\textbf{Network}} \tabularnewline
\cline{4-6}
 & & & \textbf{Seller} & \textbf{Buyer} & \textbf{Item} & & &&\tabularnewline
\hline
\hline
\parbox[t]{2mm}{\multirow{9}{*}{\rotatebox[origin=c]{90}{\hspace{-4cm} User collusion and bid-rigging}}}

&\cite{ji2007collusion}&Vickrey auction &PU &SUs & Spectrum&Buyers are divided into bidding rings presented by effective SUs. The winner is the effective SU with the highest bid. The seller sets the price for the winner to maximize the seller's expected utility gain.&Collusion resistance, and seller's utility improvement&Nash equilibrium& CRN \tabularnewline \cline{2-10}


&\cite{ji2008multi}&Double auction&PUs &SUs & Spectrum& Sellers and buyers determine their asks and bids based on CDFs. Matching each seller with each buyer is performed based on the double auction rule.&Collusion resistance, and utility improvement for both sellers and buyers&Market equilibrium&CRN  \tabularnewline \cline{2-10}

&\cite{wu2008multi} \cite{wu2009scalable}&Multi-winner auction &PU &SUs & Spectrum&Buyers are grouped into virtual bidders, and the virtual bidder with the highest bid is the winner. The payments for the winning buyers are determined so as to maximize the product of their payoffs&Collusion resistance, seller's utility and social welfare improvement&Nash bargaining equilibrium& CRN \tabularnewline \cline{2-10}

&\cite{xu2014collusion}&Double auction &Relays &Group agents & Relay service&The maximum weight matching is used to assign the winning buyers to the winning sellers. The losing buyers receive incentives extracted from the auctioneer's revenue.&Collusion resistance, budget-balance, individual rationality, truthfulness, and social welfare improvement&Market equilibrium& CWN \tabularnewline \cline{2-10}

&\cite{pan2012using}&Variant VCG auction &Licensed user &Unlicensed users & Spectrum&Buyers' bids are encrypted using the homomorphic encryption. The payment is determined based on the Shamir's polynomial secret sharing and the pricing policy of the VCG auction.&Bid-rigging resistance, spectrum utility improvement&Bayesian Nash equilibrium& CRN  \tabularnewline \cline{2-10}

&\cite{abdelhadi2015multitier}&Variant VCG auction &Broker &Base stations & Spectrum&Same as \cite{pan2012using}, but the encrypted bids are randomized by a random constant. &Bid-rigging resistance, spectrum utility improvement&Bayesian Nash equilibrium&CRN  \tabularnewline \cline{2-10}

\hline
\parbox[t]{2mm}{\multirow{9}{*}{\rotatebox[origin=c]{90}{\hspace{1cm} False-name bids}}}

&\cite{wangrobust}&General pricing &PU &SUs &Spectrum& Each buyer is assigned a price which is determined based on its critical neighbor. The auctioneer selects buyers which have bids greater than their computed prices as the winners. &False-name bid resistance, spectrum utility and revenue improvement&Optimal solution &CRN  \tabularnewline \cline{2-10}

& \cite{rathinakumar2016gavel}&Ascending-bid auction&LSA repository &LSA licensees &Spectrum& Same as \cite{zhang2013ascending}, but a buyer will be the winner if the aggregate demand of the buyer and its neighbors is less than the supply. &False-name bid cheating resistance, and privacy preserving &Walrasian
equilibrium& CN \tabularnewline \cline{2-10}
\hline
\end{tabular}
\par\end{centering}
\end{table*}

\begin{table*}[h]
\caption{A summary of advantages and disadvantages of major approaches for protecting against collusion and false-bid name cheating}
\label{table_sum_advantage_false_bid}
\scriptsize

\begin{centering}
\begin{tabular}{|>{\centering\arraybackslash}m{2cm}|>{\centering\arraybackslash}m{8.2cm}|>{\centering\arraybackslash}m{6.5cm}|}
\hline
\cellcolor{myblue} &\cellcolor{myblue} &\cellcolor{myblue} \tabularnewline
\cellcolor{myblue} \multirow{-2}{*} {\textbf{Major approaches}} &\cellcolor{myblue} \multirow{-2}{*} {\textbf{Advantages}} &\cellcolor{myblue} \multirow{-2}{*}{\textbf{Disadvantages}} \tabularnewline
\hline
\hline
\cite{ji2007collusion}&\begin{itemize} \item Always guarantee the high utility for the PU \end{itemize} & \begin{itemize}  \item Support only one PU and multiple SUs \item Support only the PU with a single channel\end{itemize}\tabularnewline \cline{2-3}
  \hline
 \cite{wu2008multi} & \begin{itemize} \item Support the PU with multiple channels  \item Avoid the collusion among the losing SUs \end{itemize} & \begin{itemize} \item Ignore sublease collusions among the winning SUs and the losing SUs \item Ignore collusion between the auctioneer and greedy SUs\end{itemize} \tabularnewline \cline{2-3}
  \hline
\cite{pan2012using} & \begin{itemize} \item Support the PU with multiple channels  \item Avoid collusion between the auctioneer and the greedy SUs\end{itemize} & \begin{itemize} \item Have high overhead communication \end{itemize} \tabularnewline \cline{2-3}
  \hline
 \cite{wangrobust}& \begin{itemize} \item Have high spectrum utilization \end{itemize} & \begin{itemize} \item Do not support the privacy preserving SUs' bids \end{itemize} \tabularnewline \cline{2-3}
\hline
\end{tabular}
\par\end{centering}
\end{table*}

\section{Miscellaneous issues}
\label{sec:Survey_security_issues}
In this section, we review some other issues related to the
use of pricing models for the wireless security.

\subsection{Spectrum Sensing Data Falsification Attack in CRNs}
\label{sec:Survey_Con_In_Au_Mess_Con}
In CRNs, Secondary Users (SUs) perform the spectrum sensing to determine if a certain channel is occupied by the Primary User's (PU's) transmission. A secondary Base Transceiver Station (BTS) then decides the spectrum allocation based on the sensing results of the SUs. However, some malicious SUs can report falsified sensing results which is commonly known as Spectrum Sensing Data Falsification (SSDF) attack. Such attack dramatically degrades the accuracy of the final decision of the BTS and the effectiveness of spectrum sensing \cite{wang2014secure}. To prevent the SSDF attack, a reputation-based sensing strategy can be adopted as proposed in \cite{lin2015secure}. Accordingly, each SU constructs its own reputation table which stores reputation values of the SU towards other SUs in the network. The values are determined based on the number of the correct/wrong sensing results that the SU has received from other SUs. The SU then combines its own sensing result with the results of other SUs which have reputation values greater than a threshold to improve the accuracy of sensing. Finally, the VCG auction was adopted to perform a distributed cheat-proof spectrum allocation. However, with the proposed solution, SUs are required to continuously update their local tables which may increase significantly energy consumption.

\subsection{Misbehaviors in Non-Cooperative Wireless Networks}
\label{sec:Survey_Con_In_Au_Mess_Non_Cooperative}
Non-cooperative wireless networks are formed by numerous heterogeneous nodes such as selfish and Byzantine nodes~\cite{su2014multi}, i.e., malicious nodes, which have different intentions and behaviors. This makes network operations such as information forwarding challenging. To handle the misbehaviors of the nodes, the authors in~\cite{su2014multi} proposed a unified framework which consists of two schemes, i.e., the Generalized Second-Price (GSP) auction~\cite{edelman2005internet} and FORBID mechanism~\cite{Su2011FORBID}. The GSP auction is to incentivize selfish nodes to cooperate while the FORBID mechanism detects malicious nodes and isolates them from packet forwarding paths. In this framework, a source as a buyer purchases the packet forwarding service from other nodes, i.e., sellers, to send data to the buyer's destination. To stimulate the selfish nodes to cooperate, each node in the forwarding path will receive a payment from the source based on its submitted ask. Then, when node $i$ in the path receives a packet from the source, it will forward this packet to the next hop node $j$ and keep an ACK message from node $j$ as a receipt. This receipt is used to get the payment from the source and identify the suspects of malicious nodes. If node $i$ cannot receive the ACK message from node $j$, node $i$ will suspect that node $j$ is malicious and record this information. At the same time, node $i$ is required to retransmit the packet with its highest transmission power such that at least one neighbor node can overhear, and this action of node $i$ can be considered to be a witness during the detection process.

Apart from the aforementioned misbehaviors, there are other issues needed to be investigated. For example, malicious nodes can collude to forge that a packet is successfully transmitted to the destination, or instead of always choosing to reject the packets, they can forward packets occasionally to deceive other nodes.

\subsection{Faked Sensing Attack in Crowdsensing Networks}
\label{sec:Survey_security_issues_faked_sensing_attack}
In crowdsensing networks, a large number of mobile devices such as smartphones and tablets are used to gather and then send their sensing data, e.g., road traffic condition, to a server. However, the users may provide faked sensing reports with the aim of saving sensing costs and avoiding privacy leakage. Therefore, pricing models can be used to motivate the users to submit true sensing reports and then to pay them according to sensing accuracy. The first-price sealed-bid reverse auction was used in \cite{xiao2015secure} to achieve the goal. In this model, the users as sellers choose asking prices, and the server as the buyer determines the payment strategy to maximize their utility functions. In particular, the utility functions of users are formed from the actual costs which are unknown by the server. The server thus employs the reinforcement learning method, i.e., Q-learning \cite{watkins1992q}, to determine its optimal payment strategy according to history of the utility received for the previous trials. Such a payment strategy can motivate the users to send high quality sensing reports while significantly decreasing the number of faked sensing reports. 


\textbf{Summary:} This section discusses the applications of pricing models to address miscellaneous security issues in wireless networks. The approaches aim to detect and prevent faults or risks caused by attackers in CRNs, non-cooperative networks, and crowdsensing networks. Also, combining auctions with learning algorithms is a useful solution to prevent the attackers. Thus more advanced learning techniques need to be investigated.

\section{Challenges, and future research directions}
\label{sec:Open_issues}
\begin{figure}[t!]
 \centering
\includegraphics[width=7.5 cm, height=4.1cm]{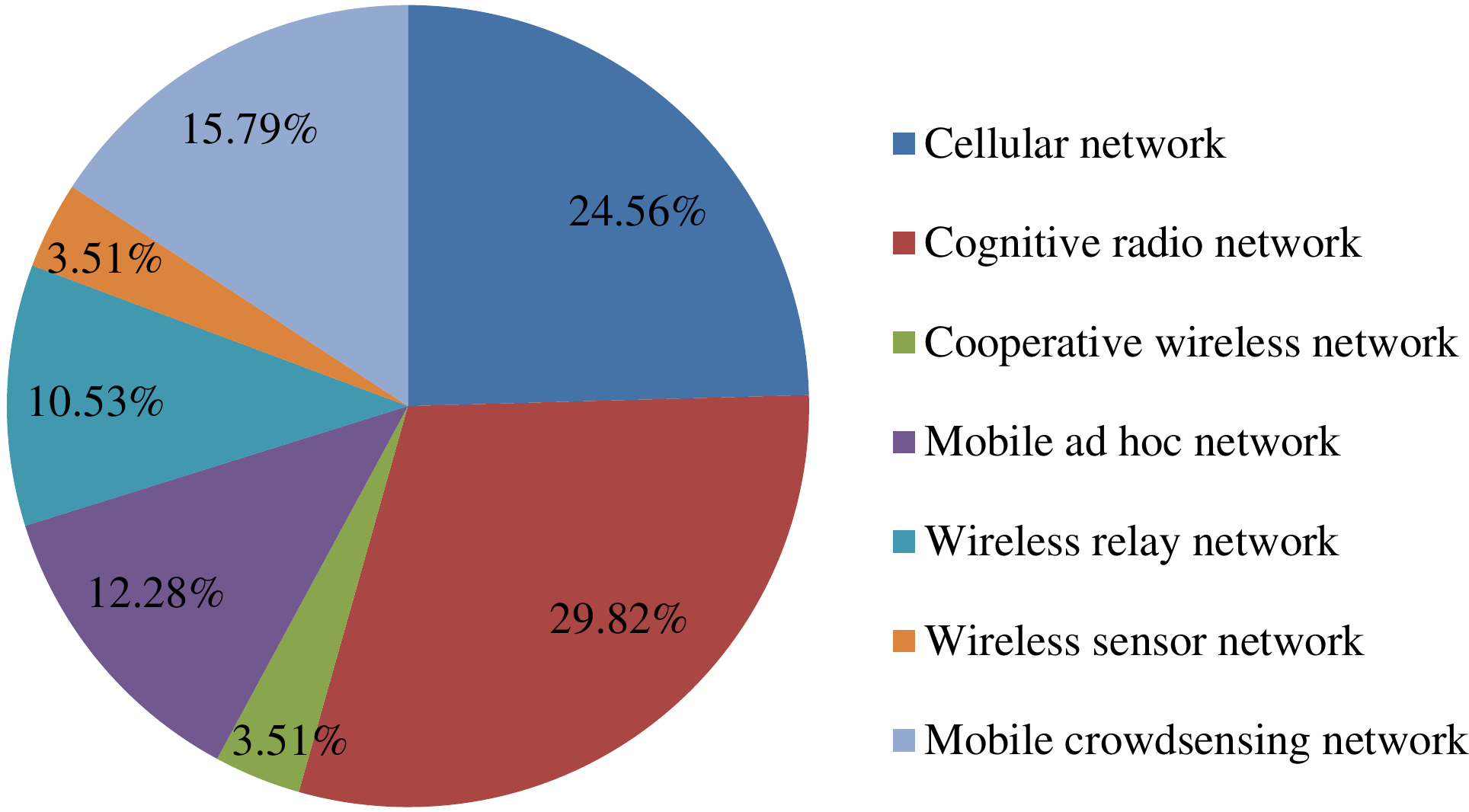}
 \caption{Summary information of percentage of wireless networks.}
 \label{Summary_Figure}
\end{figure}

A number of approaches reviewed in this survey evidently show that economic and pricing models can address diverse issues of wireless security. The ratios of economic and pricing approaches for different types of wireless networks are shown in Fig.~\ref{Summary_Figure}. From the figure, we observe that majority of the approaches are for cognitive radio networks and cellular networks while few approaches are for cooperative wireless networks and wireless sensor networks. The following discusses further some existing challenges and new research directions.

\subsection{Existing Challenges And Research Directions From Attack Perspective}
\label{sec:Open_issues_Attack_perspective}
The proposed approach in \cite{zhang2016designing} provided $k$-anonymity location privacy for the buyers with the same privacy requirements. However, the buyers practically have the different privacy requirements, i.e., different values of $k$. A possible solution is to divide the buyers into groups where buyers in a group have the same $k$-anonymity requirement. However, the challenge is how to determine the winning groups, especially when bids in the same groups are considerably different. In fact, the buyers with high bids can wait for more buyers arriving and then run another auction. However, running the auction again on the same buyers can make the auction lose the truthfulness. 

Another issue in \cite{zhang2016designing} is that the buyers and the sellers can cheat their roles. For example, a buyer may pose as a seller when participating in the auction. It might win the auction and be added into the anonymity set. In this case, the buyer receives privacy protection for free or even a payment from other buyers. Designing a mechanism to guarantee that the users will not lie their roles is an open problem.

\subsection{Research Directions From Economic Tool Perspective}
\label{sec:Open_issues_Economic_perspective}

\subsubsection{Contract theory}
\label{sec:Open_issues_Contract_Theory_Physical}
Almost all the solutions for preventing eavesdropping presented in Section~\ref{sec:Survey_eavesdropper_attack} require the support of the friendly jammer. However, the jamming cooperation is only effective if the location information of the friendly jammer is available to the source. Due to the location privacy, the friendly jammer's location might be unknown to the source. Under such an asymmetric information scenario, the contract theory \cite{li2016incentive} between the sources and the friendly jammers can be adopted. However, how to determine performance-price bundles to guarantee the highest utility for the sources while motivating the friendly jammers to cooperate needs to be developed.

\subsubsection{Cyber insurance}
\label{sec:Open_issues_Cyber_Insurance_Wireless}
Although there is a great deal of effort in developing security solutions, still it is impossible to achieve a perfect/near perfect security protection~\cite{pal2014will} because of many factors, e.g., users' behavior, network infrastructure, and a wide variety of attacks. As a result, cyber insurance is considered to be a potential and efficient solution for cyber residual risk transfer and wireless security improvement. In~\cite{jin2012spectrum}, the authors proposed an idea of using insurance in spectrum trading to improve spectrum efficiency in CRNs. In particular, the authors showed that by purchasing the PU's insurance, the SU will be insured against the potential accident, e.g., transmission failure incurred by the excessively low SINR. Meanwhile, the PU will benefit, e.g., gaining a revenue, from providing insurance services. Similarly, in~\cite{hoang2017charging}, cyber insurance was introduced as an effective solution to transfer cyber risks, e.g., unavailable information, of Plug-in-Electronic Vehicle (PEV) users to a cyber insurance company. Under the insurance coverage, a PEV user is always guaranteed the best price for charging/discharging even without information from the power gird. Clearly, cyber insurance will be one of the promising risk-mitigation solutions for wireless security. However, many open issues arise, e.g., how to determine optimal premium and insurance policy for different parties in the networks.

\subsection{Pricing Models for Security Issues in 5G HetNets}
\label{sec:Open_issues_generation_network}
Heterogeneous Networks (HetNets) are one of key technologies which will be deployed in the fifth generation (5G) network to support the deluge of data traffic. Due to the reduced cell size in HetNets, users might move through multiple small cells, e.g., picocell and femtocell, during their communication session. The security is thus more challenging due to the possible
involvement of untrusted or compromised small cells during handover \cite{duan2015authentication}. Cryptographic methods which have computation burden and complexity to both the small cells and user sides are undesirable for 5G low-power small cell infrastructures. Alternatively, economic and pricing models can provide distributed solutions which maximize the secrecy capacity for the 5G communication links without requiring additional secure channels for key exchanges as well as the complete knowledge of all channel information. 
\section{Conclusions}
\label{sec:Conclusion}
This paper has presented a comprehensive survey of the economic and pricing theory as well as their applications for wireless security issues. Firstly, we have presented an overview of the security issues in wireless networks. Then, we have introduced different economic and pricing models to understand motivations of using these models for the security issues. Afterwards, we have provided detailed reviews, analyses, and comparisons of the approaches using economic and pricing models to address a variety of security issues in wireless networks, i.e., eavesdropping attack, DoS attack, privacy
leakage, spectrum sensing data falsification attack, faked sensing attack, and illegitimate behaviors of malicious users. Finally, we have outlined open issues and future research directions. In conclusion, this paper will be a keystone for understanding how to apply the economic and pricing theories to address the security issues in wireless networks. This will be first step for further deep and broad research of the economic and pricing theories in next generation wireless networks.

\bibliographystyle{IEEEtran}
\bibliography{WirelessSecurityDatabase}{}
\end{document}